\documentclass[12pt,amssymb,superscriptaddress,preprint,letterpaper,floatfix]{revtex4-2}

\usepackage{graphicx}
\usepackage{dcolumn}
\usepackage{bm}
\usepackage{amsmath,amssymb}
\usepackage{hyperref}
\usepackage{textcomp}
\usepackage{xcolor}
\usepackage{booktabs}
\usepackage{lineno}

\makeatletter
\renewcommand{\@caption@fignum@sep}{\space}
\makeatother

\hypersetup{
  colorlinks=true,
  linkcolor=blue,
  filecolor=blue,
  urlcolor=blue,
  citecolor=blue,
}

\begin{document}

\title{Programmable memristive circuits for reconfigurable collective pseudomagnetic ordering, transport, and learning}

\author{Sichang Qiu}
\thanks{These authors contributed equally}
\affiliation{State Key Laboratory of Millimeter Waves, Southeast University, Nanjing 210096, China}
\author{Yuxin Jiang}
\thanks{These authors contributed equally}
\affiliation{Aerospace Information Research Institute, Chinese Academy of Sciences, Beijing 100094, China}
\author{Yongchao Wu}
\thanks{These authors contributed equally}
\affiliation{State Key Laboratory of Millimeter Waves, Southeast University, Nanjing 210096, China}
\author{Peng Han}
\affiliation{School of Software Engineering, Faculty of Electronic and Information Engineering, Xi'an Jiaotong University, Xi'an 710049, China}
\author{Xinxin Gao}
\affiliation{State Key Laboratory of Millimeter Waves, Southeast University, Nanjing 210096, China}
\author{Zekun Huang}
\affiliation{State Key Laboratory of Millimeter Waves, Southeast University, Nanjing 210096, China}
\author{Ce Shang}
\email{shangce@aircas.ac.cn}
\affiliation{Aerospace Information Research Institute, Chinese Academy of Sciences, Beijing 100094, China}
\author{Ronny Thomale}
\email{ronny.thomale@uni-wuerzburg.de}
\affiliation{Institute for Theoretical Physics and Astrophysics, University of W\"urzburg, Am Hubland, D-97074 W\"urzburg, Germany}
\author{Tie Jun Cui}
\email{tjcui@seu.edu.cn}
\affiliation{State Key Laboratory of Millimeter Waves, Southeast University, Nanjing 210096, China}
\author{Shuo Liu}
\email{liushuo.china@seu.edu.cn}
\affiliation{State Key Laboratory of Millimeter Waves, Southeast University, Nanjing 210096, China}
\date{\today}

\begin{abstract}
Programmable hardware typically uses fixed architectures, whereas complex physical systems obtain diverse behavior by connecting reusable local elements through different interaction rules. Memristive devices combine writable states with nonlinear dynamics, but translating these local properties into reconfigurable collective order and symmetry-governed transport remains challenging. Here we develop modular circuits configured through programmable connections between identical memristive units. Experiments establish binary conductance states as Ising-type pseudospins and show that signed feedback between units selects uniform or staggered pseudospin order. Circuit-model simulations then examine controllable domain-wall motion and pattern recovery. A symmetry encoder combines measured conductance patterns with sublattice labels and a selected symmetry rule to configure a fixed transport layer, implementing circuit analogs of ferromagnetic, antiferromagnetic, and altermagnetic transport symmetries. Although the antiferromagnetic and altermagnetic encodings share a compensated pattern, their distinct voltage spreading illustrates how programmed circuit symmetry affects transport. Device-calibrated simulations show that memristive dynamics generates history-dependent features for reservoir computing. By separating local memory from programmable interactions, symmetry encoding and task-specific readout, this architecture provides a modular route from device dynamics to collective order, symmetry-dependent transport, and learning.
\end{abstract}

\maketitle

\section{Introduction}

Complex electronic functions are commonly achieved by designing a dedicated circuit for each task. An alternative is to build on reusable local elements with electrically writable states. Programmable interactions govern their collective behavior~\cite{Strogatz2001ComplexNetworks}, while encoding and readout translate the resulting configurations into system-level functions. Memristive devices provide a natural basis for this architecture because their conductance is both electrically writable and directly measurable~\cite{Pershin2009,Yang2013,Zidan2018}. These properties allow conductance states to be updated during collective evolution and the resulting patterns to be read out for subsequent operations. The central challenge is to use collectively formed conductance patterns to program reconfigurable circuit responses.

Pseudomagnetic order, i.e., the ordering of Ising-type pseudospins encoded by binary conductance states, provides a particularly revealing setting for addressing this problem. This description connects naturally to collinear magnets in the spin-conserving limit, where $U(1)$ spin-rotation symmetry preserves the spin-up and spin-down sectors represented by these binary pseudospin labels. Ferromagnets (FMs) combine aligned local moments with finite magnetization~\cite{Igor2004,Mott1936,Fert1976,Tedrow1973,Fert2008,McGuire1975,Baibich1988}. Conventional collinear antiferromagnets (AFMs) and altermagnets (AMs) instead both contain oppositely oriented sublattices with zero net magnetization~\cite{Jungwirth2016,Baltz2018,Ramazashvili2008,Smejkal2018,Jungwirth2018}, but differ in the symmetry relations between their opposite-spin sublattices: translation- or inversion-related operations in conventional AFMs and crystal rotations in AMs~\cite{Wadley2018,Smejkal2022,Sinova2022,Jungwirth2026,Song2025,Zhou2024,Krempasky2024,Amin2024,Mazin2022}. This difference enables altermagnets to exhibit momentum-dependent spin splitting and anisotropic transport despite their magnetic compensation~\cite{GonzalezHernandez2021,Smejkal2020,Feng2022}. These distinctions motivate circuit analogs in which the same compensated pseudospin pattern supports different transport responses through distinct spatial-symmetry encodings. Their realization therefore requires separate control over the pseudospin configuration and the symmetry rule through which it enters the transport network.

Electrical networks offer direct control over the effective onsite and hopping terms that implement these symmetry rules. Their circuit Laplacians admit a tight-binding (TB) representation, with diagonal and off-diagonal entries determined by grounding and inter-node connections, linking circuit parameters to spectra and real-space voltage responses~\cite{Helbig2019,Helbig2020,Qiu2025,Jia2026,Albert2015,Ningyuan2015,Imhof2018,Lee2018,Ezawa2018,Jiang2026}. A prescribed Laplacian describes transport for a selected configuration and symmetry assignment. Linking collective ordering to transport programming additionally requires pseudospin configurations that develop through interactions and can subsequently be read out for encoding. Memristive units provide this capability through their writable and measurable conductance states. Signed feedback between units favors uniform or staggered pseudospin configurations~\cite{Onsager1944,Dowling2022}. A subsequent encoder combines the resulting conductance pattern with prescribed sublattice labels and a selected symmetry rule to determine the onsite and coupling terms of the transport network. These stages establish a modular architecture that links collective pseudospin ordering to symmetry-programmable transport.

Beyond binary pseudospin encoding, memristive conductance can evolve continuously in response to sequential inputs. Its history-dependent dynamics retain information from preceding inputs and generate nonlinear features, providing the memory and nonlinearity required for physical reservoir computing~\cite{Du2017RC,Moon2019RC,Zhong2021RC,Tanaka2019PRC,Spagnolo2022PhotonicMemristor}. The binary and continuous operating regimes thus extend the role of a common elementary unit from collective pseudospin ordering and symmetry encoding to information processing.

\begin{figure*}[h]
  \centering
   \includegraphics[width=\linewidth]{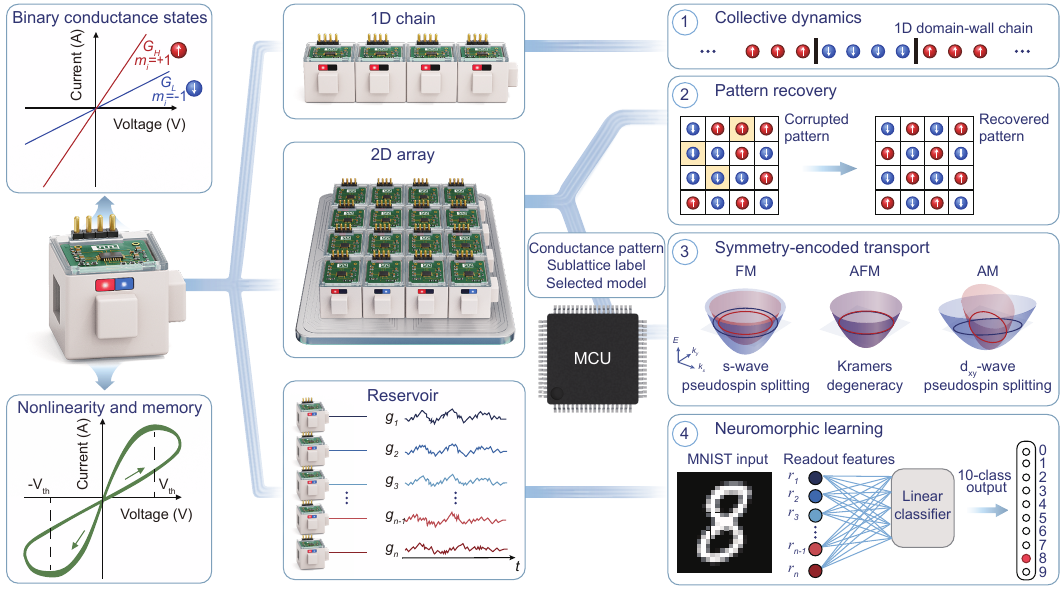}
  \caption{\textbf{Modular memristive platform for programmable collective dynamics, symmetry-encoded transport and neuromorphic learning.}
  A reusable memristive unit provides two programmable conductance states, $m_i=\pm 1$, together with nonlinear, history-dependent current–voltage characteristics. The units can be interconnected on demand to form a one-dimensional chain, a two-dimensional array or a reservoir for sequential input processing. These modular architectures support four representative functions: (1) domain-wall formation and evolution in a one-dimensional chain; (2) collective recovery of a corrupted two-dimensional pattern; (3) symmetry-encoded transport, in which a microcontroller unit (MCU) uses the measured conductance pattern, sublattice labels, and selected operating mode to program the transport network for ferromagnetic, antiferromagnetic, or altermagnetic transport analogs; and (4) neuromorphic learning, in which the time-dependent reservoir responses are used as features for linear classification of handwritten digits. Red and blue symbols denote the two binary local states.
  }
  \label{fig1}
\end{figure*}

Here we establish a modular memristive-circuit framework that combines reusable elementary units with programmable feedback, symmetry encoding, and readout (Fig.~\ref{fig1}). In the binary regime, experiments demonstrate electrically writable conductance states that encode local pseudospins and show that the polarity of nearest-neighbor (NN) feedback selects uniform or staggered configurations. Circuit-model simulations further demonstrate domain-wall depinning and recovery of corrupted patterns. A microcontroller unit (MCU) combines measured $6\times6$ conductance maps with sublattice labels and a selected FM, AFM, or AM encoding rule to program the onsite terms and diagonal couplings of a fixed transport network. Reconstructed spectra and measured voltage distributions resolve the distinct signatures of these three transport analogs in momentum and real space. In the continuous regime, device-calibrated simulations use the history-dependent responses of the memristive units to generate features for 10-class MNIST handwritten-digit classification. These results establish a modular architecture in which a common memristive unit supports collective pseudospin dynamics, symmetry-programmable transport, and reservoir computing.

\section{Feedback-assembled collective dynamics}

\subsection{Electrically writable pseudospins}

Each memristive unit encodes an electrically writable Ising-type pseudospin through its two conductance state (Fig.~\ref{fig3}a--c),
\begin{equation}
  m_i=\begin{cases}
    +1, & G_{M,i}=G_{\mathrm H},\\
    -1, & G_{M,i}=G_{\mathrm L},
  \end{cases}
  \label{eq:local_state}
\end{equation}
where $G_{M,i}=1/R_{M,i}$ is the conductance of the $i$-th memristive element. The high- and low-conductance states correspond to opposite pseudospin orientations. The resistance evolves under the applied voltage according to the threshold rule~\cite{Pershin2009,Dowling2022}
\begin{equation}
\frac{dR_{M,i}}{dt}=\begin{cases}
\operatorname{sgn}(V_i)\beta\left(|V_i|-V_t\right), & |V_i|\geq V_t,\\
0, & |V_i|<V_t,
\end{cases}
\label{eq:threshold_update}
\end{equation}
where $V_i$ is the voltage across the memristive element, $V_t$ is the switching threshold and $\beta$ is the update rate. Equation~\eqref{eq:threshold_update} makes the write--retain condition explicit: $R_{M,i}$ evolves when $|V_i|\geq V_t$ and remains unchanged when $|V_i|<V_t$. During collective operation, the external
drive and feedback jointly determine $V_i$, and hence whether the conductance state is updated or retained. Thus, $m_i$ is an electrically programmed, voltage-history-dependent circuit variable whose evolution is controlled by threshold crossings. The measured current--voltage response shows hysteretic switching between
$G_{\mathrm H}$ and $G_{\mathrm L}$ (Fig.~\ref{fig3}b,c). Cycle-resolved measurements of all 36 units further demonstrate reproducible programming and readout, with consistently separated high- and low-conductance states across the array (Supplementary Materials~\cite{suppmat}).

\subsection{Feedback-selected order}

Identical units are assembled into collective networks by programming the polarity of their NN feedback, without modifying the elementary units themselves. In each unit, the external writing signal and the currents returned from neighboring units are summed at the integrator input. The input resistance $R_{\mathrm{in}}$ and integrating capacitance $C_f$ set the characteristic time scale, whereas the coupling resistance $R_c$ sets the relative strength of the NN feedback. The nonlinear stage drives the integrator output $x_i$ towards one of two saturated branches, and the resulting branch controls the binary memristive state and provides the signal returned to neighboring units. Switching between the direct and inverter-assisted coupling paths changes the net feedback polarity.

Let $x_i$ denote a normalized continuous state derived from the integrator output of unit $i$, with $x_i=0$ chosen midway between its two saturated branches. The positive and negative branches 
program the associated memristive element into $G_{\mathrm H}$ and $G_{\mathrm L}$, respectively. The binary variable defined in Eq.~\eqref{eq:local_state} can therefore be expressed as $m_i=\operatorname{sgn}(x_i)$. With the normalized time $\tau=t/(R_{\mathrm{in}}C_f)$, the coupled dynamics are
\begin{equation}
\frac{dx_i}{d\tau}
=
u_i(\tau)-\mu x_i-\nu x_i^3
+\lambda\sum_{j\in\mathcal N_i}\xi_{ij}x_j,
\label{eq:coupled_state_main}
\end{equation}
where $u_i(\tau)$ is the normalized writing drive, $\mu$ describes leakage, $\nu>0$ provides nonlinear saturation, $\lambda$ is the normalized feedback strength and $\mathcal N_i$ denotes the set of units coupled to unit $i$. The coefficient $\xi_{ij}=\pm1$ specifies the net polarity of the feedback from unit $j$ to unit $i$. Because the signal may undergo inversion at several stages of the circuit, $\xi_{ij}$ is determined by the parity of all inversions around the complete feedback loop rather than by the coupling-path inverter alone.

To determine which spatial mode is most strongly amplified by the feedback, we set $u_i=0$ and linearize Eq.~\eqref{eq:coupled_state_main} about the unbiased fixed point $x_i=0$. For a periodic one-dimensional chain with uniform NN polarity $\xi_{ij}=\xi$, we use the Fourier-mode ansatz
\begin{equation}
x_i(\tau)\propto
\exp\left[\gamma_\xi(q)\tau\right]
\exp\left(\mathrm{i}qi\right),
\label{eq:fourier_mode_main}
\end{equation}
where $i$ labels the lattice site, $\mathrm{i}$ is the imaginary unit, and $q=2\pi\ell/N$ is the dimensionless wavevector for a chain of $N$ units, with $\ell=0,\ldots,N-1$. This gives
\begin{equation}
\gamma_\xi(q)
=
-\mu+2\lambda\xi\cos q,
\label{eq:chain_growth_rate}
\end{equation}
For $\xi=+1$, the largest growth rate occurs at $q=0$, so the feedback selects a uniform spatial mode. For $\xi=-1$ in a periodic chain with even $N$, the maximum shifts to $q=\pi$, selecting a staggered mode. When the maximum growth rate is positive, the selected mode grows until the cubic term $-\nu x_i^3$ limits the amplitude, driving the system toward a binary pattern with uniform or staggered character.

The feedback polarity therefore reproduces the ordering preference associated with the sign of an Ising interaction: $\xi=+1$ favors equal neighboring states, $m_i=m_j$, whereas $\xi=-1$ favors opposite neighboring states, $m_i=-m_j$. This correspondence concerns the selected spatial order rather than the microscopic dynamics. The circuit evolves deterministically under electrical driving, whereas Glauber dynamics describes stochastic, thermally activated spin transitions~\cite{Glauber1963}.

\begin{figure*}[h]
  \centering
  \includegraphics[width=\linewidth]{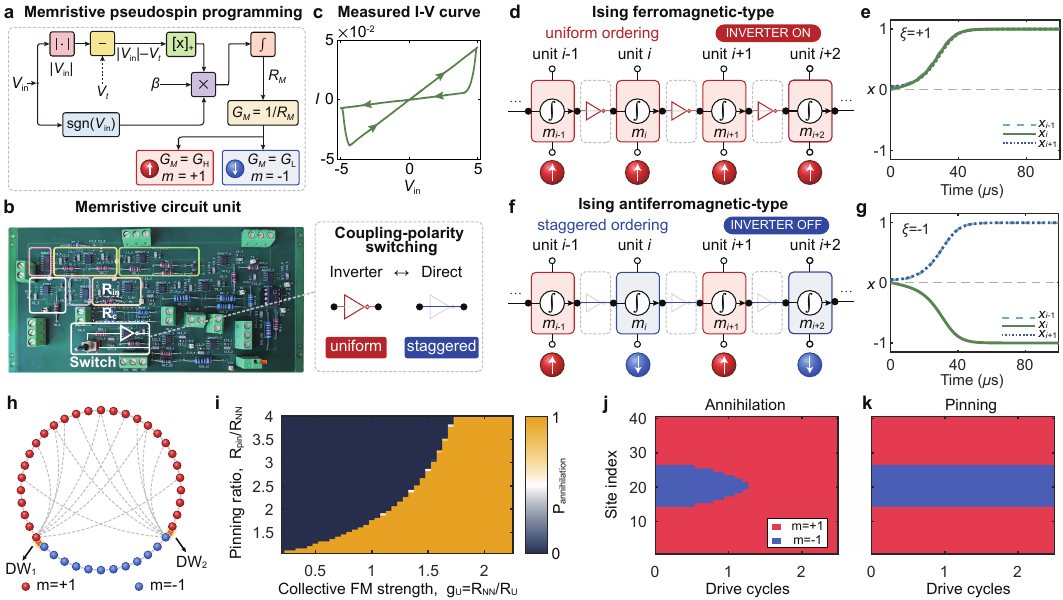}
  \caption{\textbf{Memristive pseudospin programming, feedback selection and domain-wall dynamics.} 
  \textbf{a}, Threshold-controlled programming of a memristive unit, in which the history-dependent resistance $R_M$ defines two conductance states, $G_H$ and $G_L$, corresponding to $m=+1$ and $-1$, respectively. 
  \textbf{b}, Photograph of the fabricated circuit unit. 
  \textbf{c}, Measured current--voltage hysteresis.
  \textbf{d,e}, Inverter-assisted NN feedback ($\xi=+1$) favors parallel neighboring pseudospins, driving coupled units toward a uniform pattern.  
  \textbf{f,g}, Direct NN feedback ($\xi=-1$) favors antiparallel neighboring pseudospins, producing a staggered pattern.
  \textbf{h}, Periodic chain initialized with a minority domain bounded by two pinned domain walls, $\mathrm{DW}_1$ and $\mathrm{DW}_2$.
  \textbf{i}, Domain-wall annihilation probability as a function of the collective FM-type feedback strength $g_U=R_{\mathrm{NN}}/R_U$ and the pinning ratio $R_{\mathrm{pin}}/R_{\mathrm{NN}}$. Each pixel represents the fraction of 200 independently noise-perturbed initial conditions classified as annihilated.
  \textbf{j,k}, Representative evolution showing domain-wall annihilation and pinning, respectively.}
  \label{fig3}
\end{figure*}

Using the component values of the experimental circuit,
$R_{\mathrm{in}}=1~\mathrm{k}\Omega$ and $R_c=8~\mathrm{k}\Omega$, each NN feedback branch has the
normalized strength $\lambda=R_{\mathrm{in}}/R_c=0.125$.
The inverter-assisted and direct coupling paths implement the net loop polarities $\xi=+1$ and $\xi=-1$, respectively, as illustrated in Fig.~\ref{fig3}d,f. Numerical integration of Eq.~\eqref{eq:coupled_state_main} using these circuit parameters shows that the $\xi=+1$ configuration amplifies the $q=0$ mode and evolves towards a uniform pattern (Fig.~\ref{fig3}e), whereas the $\xi=-1$ configuration amplifies the $q=\pi$ mode and evolves
towards a staggered pattern (Fig.~\ref{fig3}g).

Coupled-unit measurements confirm these local ordering rules: inverter-assisted feedback produces identical neighboring readout states, whereas direct feedback produces alternating low and high states. Representative waveforms and additional circuit measurements are provided in the Supplementary Materials~\cite{suppmat}, while the Fourier-mode derivation and mapping to Eq.~\eqref{eq:coupled_state_main} are detailed in Methods.

\subsection{Domain-wall depinning and annihilation}

Having established feedback-selected bulk order, we next use a finite minority domain to examine how local bond programming and collective feedback jointly control a localized defect. We consider a periodic chain of $N=40$ memristive units initialized with a 12-site minority domain, whose two interfaces define a pair of domain walls (Fig.~\ref{fig3}h). All NN coupling branches have resistance $R_{\mathrm{NN}}=8~\mathrm{k}\Omega$, except for the two bonds initially located at the domain interfaces, whose resistance is set to $R_{\mathrm{pin}}=r_{\mathrm{pin}}R_{\mathrm{NN}}$. A domain wall moves when a unit adjacent to the interface switches to the opposite state, shifting the interface by one lattice site. Because the NN feedback favors equal neighboring states, the two neighbors exert competing influences on a boundary unit: the opposite-state neighbor across the interface promotes reversal, whereas the same-state neighbor within the domain stabilizes its current state. Since the feedback strength is inversely proportional to the branch resistance, setting $r_{\mathrm{pin}}>1$ selectively weakens the reversal-promoting feedback across the interface. The same-domain feedback then dominates, suppressing boundary-unit switching and pinning each wall near its corresponding programmed bond.

To drive the pinned minority domain toward the surrounding background state, we introduce a self-excluded collective FM-type feedback channel,
\begin{equation}
    I_i^{(U)}
    =
    -\frac{1}{R_U}\frac{1}{N}\sum_{j\neq i}x_j
    =
    -\frac{g_U}{R_{\mathrm{NN}}}\frac{1}{N}\sum_{j\neq i}x_j,
    \qquad
    g_U\equiv\frac{R_{\mathrm{NN}}}{R_U}.
    \label{eq:collective_feedback}
\end{equation}
For the configuration considered here, the summed signal is dominated by the majority background state and therefore drives the minority domain toward it, promoting inward wall motion and annihilation. Excluding $j=i$ removes self-feedback, while $1/N$ normalizes the signal with system size. The minus sign follows from the current convention of the inverting summing stage; the complete loop still favors equal-state alignment. Thus, increasing $g_U$ promotes depinning and annihilation, whereas increasing $r_{\mathrm{pin}}$ favors pinning.

For each parameter pair $(g_U,r_{\mathrm{pin}})$, we simulate 200 trajectories initialized from the same prewritten minority-domain configuration with independent voltage perturbations. The binary configuration is sampled once per drive cycle. A trajectory is classified as annihilated when no domain wall is detected for three consecutive stroboscopic samples and the final magnetization $M=1/N\sum_{i=1}^{N}m_i$
satisfies $|M|\geq0.9$. 

Figure~\ref{fig3}i shows the empirical annihilation probability, with each pixel representing the fraction of the 200 trajectories classified as annihilated. 
As $r_{\mathrm{pin}}$ increases, this crossover shifts towards larger $g_U$, showing that
a larger collective feedback strength is required to overcome the more strongly weakened interface bonds and eliminate the minority domain. Representative annihilated and pinned trajectories are shown
in Fig.~\ref{fig3}j and Fig.~\ref{fig3}k, respectively. Additional simulation parameters and convergence tests are provided in the Supplementary Materials~\cite{suppmat}.

\section{Competition between staggered and uniform feedback: state selection and recovery}
In a two-dimensional network, we next examine whether the same competing feedback channels can both select between uniform and staggered configurations and restore either configuration after pseudospin errors.
We combine fixed NN negative feedback, which favors staggered order, with a tunable self-excluded collective positive-feedback channel. The latter corresponds to a fully connected, $1/N$-normalized Husimi--Temperley interaction~\cite{Rikvold2016} and favors uniform order. Before output saturation, the circuit dynamics are
\begin{equation}
C\dot V_i=-I_{\mathrm{drive}}(t)-\sum_{j\in\mathrm{NN}(i)}G_{ij}V_j
+\frac{1}{R_U N}\sum_{j\ne i}V_j,
\label{eq:competition_circuit_dynamics}
\end{equation}
where $R_U$ controls the strength of the collective feedback while the NN feedback network remains unchanged.

The competition between the two feedback channels can be expressed in terms of $m_i=\pm1$ as
\begin{equation}
\mathcal H_{\mathrm{comp}}
=-J_{\mathrm{NN}}\sum_{\langle i,j\rangle}m_i m_j
-\frac{J_U}{2N}\sum_i\sum_{j\ne i}m_i m_j-h\sum_i m_i,
\label{eq:competition_spin_mapping}
\end{equation}
where $J_{\mathrm{NN}}<0$ represents the staggered NN preference, $J_U>0$ represents the collective uniform preference and $h$ denotes the uniform bias associated with the common drive. Using $\sum_i\sum_{j\ne i}m_i m_j
= \left(\sum_i m_i\right)^2-N$, the collective term reduces, up to a constant, to $-J_U(\sum_i m_i)^2/(2N)$, confirming its Husimi--Temperley form. Equation~\eqref{eq:competition_spin_mapping} identifies the ordering preferences of the two feedback channels; the actual state evolution remains governed by the driven nonlinear circuit equation~\eqref{eq:competition_circuit_dynamics}. Initial conditions are sampled so that their overlap distributions with the uniform and staggered targets are statistically balanced, preventing either target class from being favored by initialization. The sampling and state-classification procedures are described in Methods.

\begin{figure*}[h]
\centering
\includegraphics[width=0.9\linewidth]{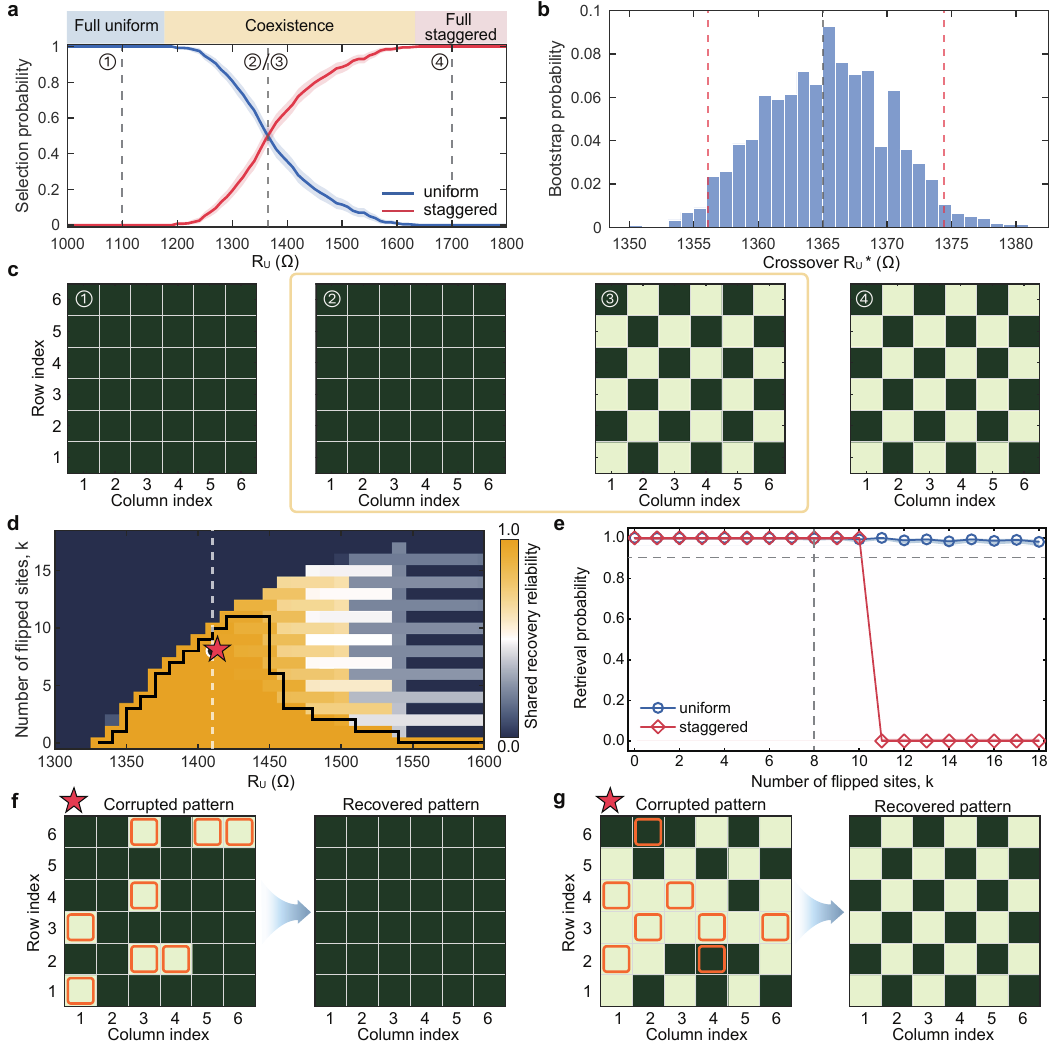}
\caption{\textbf{Feedback-controlled state selection and pattern recovery in a 2D circuit model.}
Each probability is estimated from 400 independently initialized realizations.
\textbf{a}, Probabilities of reaching the uniform and staggered states versus $R_U$; shading denotes bootstrap uncertainty.
\textbf{b}, Bootstrap distribution of the crossover resistance $R_U^*$; dashed lines indicate the median and $95\%$ confidence interval.
\textbf{c}, Representative $6\times6$ final states at the conditions marked in \textbf{a}.
\textbf{d}, Shared exact-recovery reliability versus $R_U$ and the number of flipped sites $k$. Exact recovery requires $Q=1$, and the shared reliability is the smaller of the $95\%$ Wilson lower bounds for the two target patterns. The black boundary encloses the contiguous region with shared reliability above $0.90$.
\textbf{e}, Exact-recovery probabilities at $R_U=1410~\Omega$. Horizontal and vertical dashed lines mark $0.90$ and $k=8$, respectively; error bars show $95\%$ Wilson confidence intervals and may be smaller than the symbols.
\textbf{f,g}, Recovery of the uniform and staggered targets from inputs with eight flipped sites.}
\label{fig5}
\end{figure*}

For each value of $R_U$, we perform 400 simulations with independently sampled initial conditions and determine the probabilities of reaching the uniform and staggered states. At small $R_U$, the strong collective channel predominantly selects the uniform state. Increasing $R_U$ weakens this contribution, allowing the fixed negative NN feedback to select the staggered state (Fig.~\ref{fig5}a). Bootstrap resampling gives a median crossover resistance of $R_U^*\simeq1365~\Omega$, where the two states have equal selection probabilities (Fig.~\ref{fig5}b). Its displacement from the homogeneous linear-mode estimate, $R_U^{(0)}=1000~\Omega$, results from the additional uniform-mode contribution of the common periodic drive. Near the crossover, either collective pattern can emerge under the same circuit parameters, depending on the initial state (Fig.~\ref{fig5}c).

The crossover regime also enables the same feedback setting to return to either selected collective pattern after pseudospin perturbations. To generate a corrupted input, we start from an uncorrupted uniform or staggered target, select $k$ distinct lattice sites uniformly without replacement, and reverse $m_i$ at each selected site. A trajectory is counted as successfully recovered only when its final configuration exactly matches the original uncorrupted target, corresponding to $Q=1$. For each $(R_U,k)$, we separately calculate the uniform- and staggered-target recovery probabilities and define the shared recovery reliability as the smaller of their $95\%$ Wilson lower confidence bounds. Figure~\ref{fig5}d identifies a contiguous operating region in which this shared reliability exceeds $0.90$. At $R_U=1410~\Omega$, both target patterns remain reliably recoverable after eight site flips (Fig.~\ref{fig5}e--g). These results show that competition between local staggered feedback and collective uniform feedback can be tuned to either of two selected collective state patterns after local perturbations before symmetry encoding.

\section{Symmetry encoding of FM-, AFM- and AM-type transport}
The measured binary conductance pattern defines a pseudospin configuration, but does not by itself determine the symmetry of the transport network. A uniform pattern has nonzero net pseudospin polarization and is used to encode the FM-type configuration. A compensated staggered pattern, by contrast, can encode either an AFM- or an AM-type configuration, depending on the spatial relation assigned to the hopping environments of the two sublattices. The AFM rule assigns equivalent diagonal-hopping environments to the two sublattices, whereas the AM rule relates their diagonal-hopping environments by a $90^\circ$ crystal rotation. 

We implement this distinction using an MCU-based symmetry encoder. Here, $m_i=\pm1$ denotes the pseudospin orientation encoded by the measured conductance of the $i$-th memristive unit, $\eta_i=+1$ ($-1$) labels the $A$ ($B$) sublattice, and $\chi\in\{\mathrm{FM},\mathrm{AFM},\mathrm{AM}\}$ denotes the selected transport rule. From $(m_i,\eta_i,\chi)$, the MCU calculates two binary control variables: $P_i^\chi$, which configures the grounding branch, and $Q_i^\chi$, which selects the orientation of the strong and weak diagonal couplings. The measured conductance pattern and the independently selected symmetry rule are thereby converted into switch-control instructions for the common transport layer. The programmed configurations are characterized experimentally through multiport measurements that reconstruct their momentum-space circuit-Laplacian
spectra and through local-excitation measurements that resolve their real-space voltage-spreading responses.

\subsection{Unified tight-binding description}

To elucidate how the symmetry encoder configures distinct transport analogs for the selected FM, AFM and AM operating modes, we formulate the three configurations within a unified four-state TB Hamiltonian in the basis $\Psi_{\mathbf k}=(|A,\uparrow\rangle,|B,\uparrow\rangle,|A,\downarrow\rangle,|B,\downarrow\rangle)^T$,
\begin{equation}
H_\chi(\mathbf k)=H_0(\mathbf k)+d(\mathbf k)\Gamma_\chi-\Delta\Lambda_\chi
+\zeta_\chi t_\perp\sigma_x\tau_0-\delta\sigma_z\tau_z,
\label{eq:model}
\end{equation}
where $\chi=\mathrm{FM},\mathrm{AFM},\mathrm{AM}$, $\sigma_i$ and $\tau_i$ act in the pseudospin and sublattice spaces, respectively. The common contribution is
\begin{equation}
H_0(\mathbf k)=\epsilon(\mathbf k)\sigma_0\tau_0+\gamma(\mathbf k)\sigma_0\tau_x,
\label{eq:H0}
\end{equation}
with
\begin{align}
\epsilon(\mathbf k)&=2(t_1+t_2)\cos k_x\cos k_y,\\
d(\mathbf k)&=-2(t_1-t_2)\sin k_x\sin k_y,\\
\gamma(\mathbf k)&=2t_{\mathrm{nn}}(\cos k_x+\cos k_y).
\end{align}
The order-dependent coefficients are $(\Gamma_\chi,\Lambda_\chi,\zeta_\chi)=(\sigma_0\tau_0,\sigma_z\tau_0,0)$ for the FM model, $(\sigma_0\tau_0,\sigma_z\tau_z,1)$ for the AFM model and $(\sigma_0\tau_z,\sigma_z\tau_z,1)$ for the AM model. Here $t_1$ and $t_2$ are diagonal hopping amplitudes, $t_{\mathrm{nn}}$ is the NN hopping, and $t_\perp$ weakly hybridizes the pseudospin sectors without mixing sublattices. The parameter $\Delta$ sets the dominant pseudospin offset. For the AFM and AM configurations, the onsite terms combine into $-M\sigma_z\tau_z$, where $M=\Delta+\delta$. Their explicit Hamiltonians are given in Methods.

For the FM analog, the sublattice-independent term
$-\Delta\sigma_z\tau_0$ separates the two pseudospin sectors. Provided that $\Delta>|\delta|$, the low-energy branches on both sublattices have the same pseudospin, yielding a finite pseudospin polarization. The $d(\mathbf k)\sigma_0\tau_0$ term imposes the same diagonal-hopping imbalance on both sublattices. The corresponding bands and constant-energy contours (CECs) therefore retain a common pseudospin character and directional preference (Fig.~\ref{fig2}a--c).

For the AFM analog, the staggered onsite term produces two compensated low-energy sectors. Because $d(\mathbf{k})$ is sublattice independent, these sectors are related at the same momentum by
\begin{equation}
\Theta_{\mathrm{AFM}}
=
\sigma_x\tau_x\mathcal K,
\qquad
\Theta_{\mathrm{AFM}}
H_{\mathrm{AFM}}(\mathbf{k})
\Theta_{\mathrm{AFM}}^{-1}
=
H_{\mathrm{AFM}}(\mathbf{k}),
\qquad
\Theta_{\mathrm{AFM}}^2=+1.
\label{eq:AFM_symmetry}
\end{equation}
Thus, $\Theta_{\mathrm{AFM}}$ relates the two sectors but does not itself enforce their degeneracy. Along $X$--$M$--$Y$, $\gamma(\mathbf{k})=0$, so the NN sublattice coupling vanishes and $\tau_z$ is conserved. The resulting same-momentum antiunitary symmetry is
\begin{equation}
\widetilde{\Theta}_{\mathrm{AFM}}
=
\tau_z\Theta_{\mathrm{AFM}}
=
i\sigma_x\tau_y\mathcal K,
\qquad
\widetilde{\Theta}_{\mathrm{AFM}}^{\,2}=-1,
\label{eq:AFM_boundary_antiunitary}
\end{equation}
which protects the exact twofold degeneracy along $X$--$M$--$Y$. Away from this boundary, the symmetry is lost and the AFM branches
are weakly split (Fig.~\ref{fig2}e--g). The reconstructed spectrum is shown in Fig.~\ref{fig2}h, with further details provided in Methods.

For the AM analog, the two sublattices remain compensated, but the term
$d(\mathbf{k})\sigma_0\tau_z$ produces opposite diagonal-hopping imbalances on the two sublattices. As the internal antiunitary transformation exchanges both the two effective pseudospin sectors and the two sublattices, it has the same matrix representation as in the AFM case, $\Theta_{\rm AM}=\sigma_x\tau_x\mathcal K$. The corresponding circuit symmetry operations are nevertheless different. In the AFM configuration, the internal transformation maps the Hamiltonian back to itself at the same momentum. In the AM configuration, it must be accompanied by the spatial rotation $C_{4z}$, which transforms the momentum as $(k_x,k_y)\mapsto(-k_y,k_x)$. The resulting AM symmetry relation is
\begin{equation}
\Theta_{\mathrm{AM}}
H_{\mathrm{AM}}(\mathbf k)
\Theta_{\mathrm{AM}}^{-1}
=
H_{\mathrm{AM}}(C_{4z}\mathbf k).
\label{eq:AM_symmetry}
\end{equation}
Because $d(C_{4z}\mathbf{k})=-d(\mathbf{k})$, the opposite-pseudospin branches are generally split at the same generic momentum but satisfy $E_{+}(\mathbf k)=E_{-}(C_{4z}\mathbf k)$.
Their dispersions and CECs are therefore mapped onto each other by a $90^\circ$ momentum rotation, producing the characteristic AM-type spectral structure shown in Fig.~\ref{fig2}i--k. The corresponding experimentally reconstructed spectrum is shown in Fig.~\ref{fig2}l.

\subsection{Circuit encoding of antiunitary symmetry}

The symmetry relations introduced above are implemented by transforming the pseudospin, sublattice, and link-direction labels. The circuit analog of time reversal acts on the pseudospin as $\mathcal T_m:m_i\mapsto-m_i$.
To distinguish the two diagonal directions, we assign $\rho=+1$ to the $\langle11\rangle$ direction and $\rho=-1$ to the $\langle1\bar{1}\rangle$ direction. The corresponding diagonal conductance is encoded as
\begin{equation}
G_i^\rho
=
\bar G+\frac{\rho Q_i^\chi}{2}\Delta G,
\label{eq:diag_rule}
\end{equation}
where reversing the sign of $Q_i^\chi$ exchanges the strong and weak diagonal couplings. The grounding and diagonal-coupling controls, $P_i^\chi$ and $Q_i^\chi$, are programmed according to the selected FM, AFM, or AM symmetry.

\begin{figure*}[h]
  \centering
  \includegraphics[width=\linewidth]{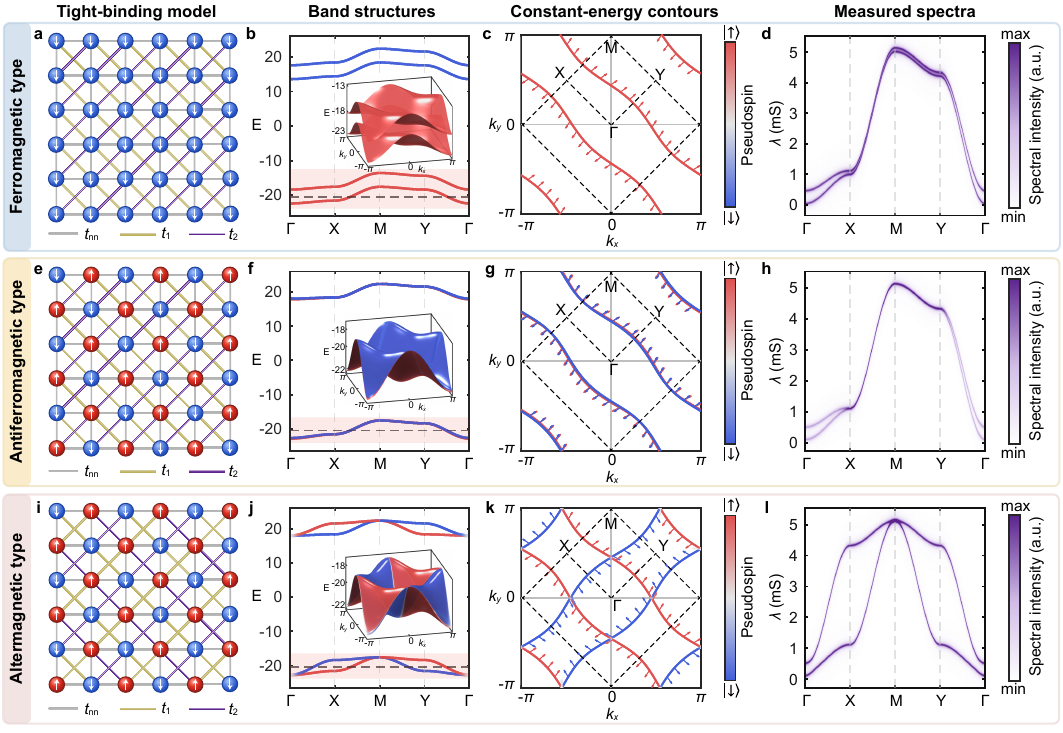}
  \caption{\textbf{Theoretical and experimentally reconstructed spectra of FM-, AFM-, and AM-type circuit configurations.} The top, middle, and bottom rows correspond to the FM-, AFM-, and AM-type configurations, respectively. The TB calculations are shown in \textbf{a-c, e-g} and \textbf{i-k}; \textbf{d,h} and \textbf{l} show spectra reconstructed from  experimentally measured multiport circuit responses.
  \textbf{a,e,i}, Tight-binding lattices showing the programmed pseudospin configurations and couplings.
  \textbf{b,f,j}, Calculated band structures along the high-symmetry path $\Gamma$–$X$–$M$–$Y$–$\Gamma$. The shaded regions identify the low-energy sectors enlarged in the insets, which show the corresponding band surfaces in momentum space; horizontal dashed lines mark the energies used for the constant-energy contours (CECs).
  \textbf{c,g,k}, CECs within the first Brillouin zone (BZ). Red and blue denote the two opposite pseudospin sectors, and dashed lines indicate the folded Brillouin-zone boundaries.
  \textbf{d,h,l}, Circuit-Laplacian spectra $\lambda(\mathbf{k})$ experimentally reconstructed from the measured multiport impedance matrices. The color scale represents spectral intensity. The FM-type response is pseudospin-polarized, whereas the AFM-type response exhibits twofold degeneracy along the $X$–$M$–$Y$ boundary. The AM-type response instead shows momentum-dependent pseudospin splitting, with the two sectors related by a $C_{4z}$ rotation.}
  \label{fig2}
\end{figure*}

For the AFM encoding, either $t_{\boldsymbol{\delta}}\mathcal T_m$ or $\mathcal P\mathcal T_m$ transforms the labels as $(m_i,\eta_i,\rho)\mapsto(-m_i,-\eta_i,\rho)$, where $\eta_i=\pm1$ labels the two sublattices. We therefore choose
\begin{equation}
Q_i^{\mathrm{AFM}}=m_i\eta_i,
\qquad
P_i^{\mathrm{AFM}}=m_i\eta_i.
\end{equation}
Because both $m_i$ and $\eta_i$ change sign, their product remains unchanged. Consequently, $P_i^{\mathrm{AFM}}$ and $\rho Q_i^{\mathrm{AFM}}$, and hence the encoded grounding and diagonal conductances, are invariant under the combined symmetry.

For the AM rule, the combined operation $C_{4z}\mathcal T_m$ reverses the binary state and exchanges both the sublattices and the diagonal directions: $(m_i,\eta_i,\rho)\mapsto(-m_i,-\eta_i,-\rho)$. We therefore choose
\begin{equation}
Q_i^{\mathrm{AM}}
=
m_i,
\qquad
P_i^{\mathrm{AM}}
=
m_i\eta_i.
\end{equation}
for which $P_i^{\mathrm{AM}}$ remains unchanged and
$\rho Q_i^{\mathrm{AM}}\mapsto(-\rho)(-m_i)=\rho Q_i^{\mathrm{AM}}$.

The FM rule starts from a uniform state $m_i=m_0$ and uses
\begin{equation}
Q_i^{\mathrm{FM}}
=
m_0,
\qquad
P_i^{\mathrm{FM}}
=
m_0\eta_i.
\end{equation}
The state reversal $\mathcal T_m$ changes the resulting circuit configuration, and no sublattice-exchanging spatial operation restores it, consistent with broken effective time-reversal symmetry.

For the momentum-space characterization, each programmed transport circuit realizes the corresponding low-energy two-band sector of the unified model. Its finite-network response is measured by sequentially exciting all 36 nodes and recording the complete voltage distribution for each source position. The resulting multiport impedance matrix is inverted to obtain the finite-circuit
Laplacian, whose spatially averaged matrix elements yield the effective $2\times2$ Bloch circuit Laplacian. The experimentally reconstructed branches reproduce the characteristic FM-, AFM- and AM-type spectral relations along the $\Gamma$--$X$--$M$--$Y$--$\Gamma$ path
(Figs.~\ref{fig2}d, h and l). Details of the multiport measurements and Bloch reconstruction are provided in Methods.

\subsection{Symmetry-dependent voltage spreading}

\begin{figure*}[h]
  \centering
  \includegraphics[width=\linewidth]{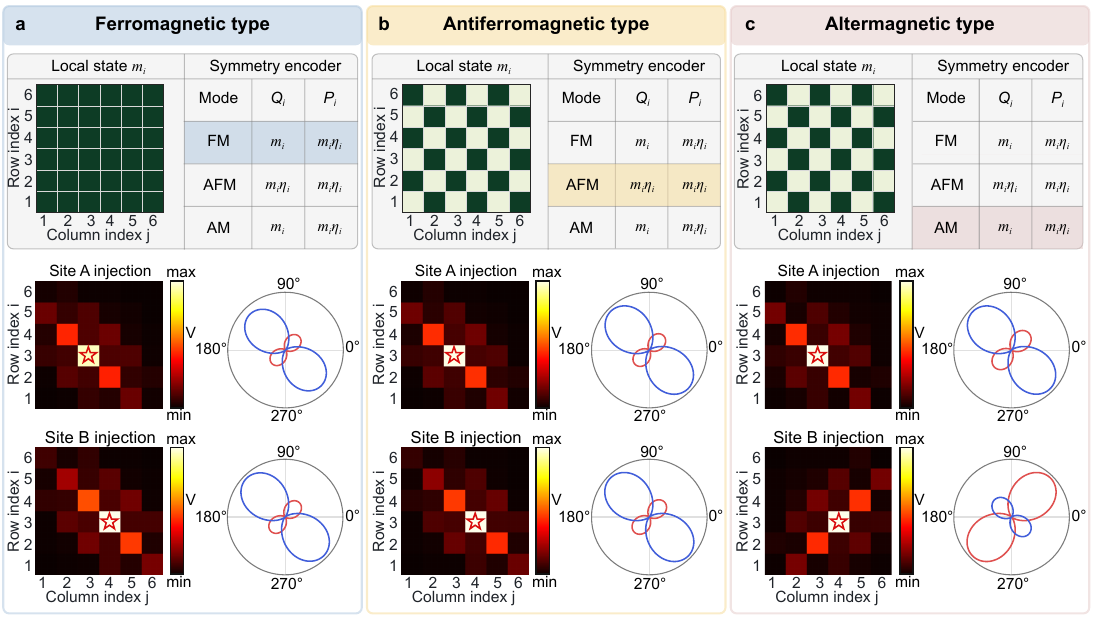}
  \caption{\textbf{Symmetry-dependent voltage spreading in a programmable transport layer.} 
\textbf{a-c}, FM-, AFM- and AM-type configurations programmed sequentially in the same $6\times6$ transport layer. The upper panels show the pseudospin configurations obtained from measured conductance patterns loaded into the MCU-based encoder, together with the selected rules for the diagonal-coupling selector $Q_i$ and grounding selector $P_i$, where $m_i$ denotes the local pseudospin orientation and $\eta_i$ labels the $A$ and $B$ sublattices. The middle and lower panels show the measured steady-state voltage distributions and their directional polar plots for excitation at representative $A$- and $B$-sublattice sites, respectively. Stars mark the injection sites, and the polar plots resolve voltage spreading along the $\langle11\rangle$ and $\langle1\bar{1}\rangle$ directions.
\textbf{a}, In the FM-type configuration, the uniform state pattern produces a common diagonal-coupling orientation but different grounding environments on the two sublattices, giving voltage distributions with the same preferred direction.
\textbf{b}, In the AFM-type configuration, the staggered pattern is encoded into equivalent transport environments on the two sublattices, again producing the same preferred direction.
\textbf{c}, The AM-type configuration uses the same staggered pattern, but encodes diagonal-coupling orientations related by a $90^\circ$ rotation, producing orthogonal preferred directions.
}
  \label{fig4}
\end{figure*}

The reconstructed Bloch spectra characterize the programmed transport circuits in momentum space. We further examine how the same symmetry relations manifest in real space by measuring voltage spreading under local excitation. The FM-, AFM-, and AM-type configurations are programmed sequentially in the same $6\times6$ transport layer using the selector rules summarized in Fig.~\ref{fig4}. For each configuration, a voltage is applied to representative sites on the two sublattices, and the steady-state voltages at all lattice nodes are recorded (Figs.~\ref{fig4}a-c). The resulting spatial profiles and their directional components along $\langle11\rangle$ and $\langle1\bar{1}\rangle$ reveal how the transport environments of the two sublattices are related.

For the FM rule, the uniform pseudospin configuration assigns the same diagonal-coupling orientation to both sublattices but different grounding configurations to the $A$ and $B$ sites. Consequently, excitation at neighboring $A$- and $B$-sublattice sites produces voltage profiles with different amplitudes but the same preferred spreading direction (Fig.~\ref{fig4}a).

The same compensated staggered pattern produces distinct AFM- and AM-type responses under different symmetry rules. In the AFM-type configuration, the two sublattices have equivalent grounding and diagonal-coupling environments, yielding similar voltage profiles with the same preferred direction (Fig.~\ref{fig4}b). In the AM-type configuration, their grounding environments remain equivalent, whereas the diagonal-coupling orientations are related by $C_{4z}$. The resulting voltage profiles exhibit preferred directions separated by approximately $90^\circ$ (Fig.~\ref{fig4}c) in the programmed circuit.

These measurements show that a compensated staggered pattern does not uniquely determine the transport response. The AFM- and AM-type responses are distinguished by the spatial relation between the sublattice hopping environments, which is imposed by the symmetry encoder. Changing the conductance pattern and the selected transport mode therefore reconfigures the real-space transport response within the same physical transport layer.

\section{Device-calibrated simulation of memristive reservoir computing}

\begin{figure*}[h]
  \centering
  \includegraphics[width=\linewidth]{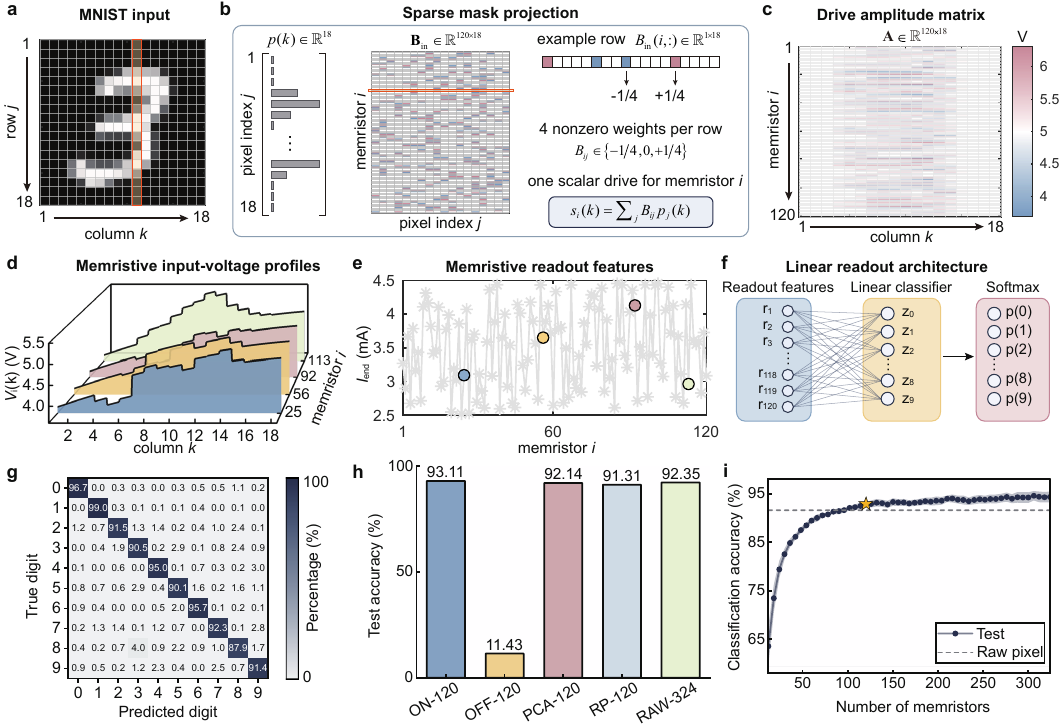}
\caption{
\textbf{Device-calibrated numerical assessment of sequence-dependent feature encoding. }
\textbf{a}, An MNIST image resized to $18\times18$ pixels and presented as a sequence of 18 columns.
\textbf{b}, Sparse projection of each image column $\mathbf{p}(k)$ onto 120 simulated memristive units using a fixed mask $\mathbf{B}_{\mathrm{in}}\in\mathbb{R}^{120\times18}$, with four nonzero weights ($\pm1/4$) per row.
\textbf{c}, Resulting drive-amplitude matrix $\mathbf{A}\in\mathbb{R}^{120\times18}$ used in the numerical reservoir.
\textbf{d}, Representative input-voltage sequences for four units.
\textbf{e}, Final simulated read currents forming the 120-dimensional feature vector.
\textbf{f}, Offline linear softmax readout used to map the 120-dimensional reservoir features onto the ten digit classes.
\textbf{g}, Row-normalized confusion matrix on the MNIST test set.
\textbf{h}, Mean test accuracies over ten independent numerical realizations for the 120-unit reservoir with state evolution enabled (ON-120) or disabled (OFF-120), compared with 120-dimensional principal-component (PCA-120) and random-projection (RP-120) representations and all 324 resized pixels (RAW-324). The PCA basis is fitted using the training data, whereas the RP matrix is randomly generated and fixed. All methods use the same class of linear readout.
\textbf{i}, Test accuracy versus reservoir size in the numerical model, averaged over ten independent realizations. The shaded band denotes one standard deviation, the star marks the 120-unit configuration used in \textbf{b--h}, and the dashed line indicates the RAW-324 baseline.
}
  \label{fig6}
\end{figure*}

The preceding results use binary conductance states to encode pseudospins for collective ordering and symmetry-programmable transport. We next use the continuous conductance dynamics of the same elementary unit to generate sequence-dependent features. The numerical reservoir comprises 120 independent instances of the device-calibrated memristive model and is evaluated on 10-class MNIST handwritten-digit classification.

Each MNIST image is resized to $18\times18$ pixels and presented as a sequence of 18 columns (Fig.~\ref{fig6}a). At each step, a fixed sparse mask $\mathbf{B}_{\rm in}$ projects the current column onto 120 drive amplitudes (Figs.~\ref{fig6}b,c), producing distinct voltage sequences and conductance trajectories across the simulated units (Fig.~\ref{fig6}d). After all columns have been applied, the final currents form a 120-dimensional feature vector (Fig.~\ref{fig6}e). The mask and device-model parameters remain fixed within each realization, and the resulting feature vectors are evaluated using an offline linear softmax readout (Fig.~\ref{fig6}f).

The confusion matrix shows discrimination across all ten digit classes (Fig.~\ref{fig6}g). The 120-unit numerical reservoir achieves a mean test accuracy of $93.11\%$ over ten independent realizations, each evaluated on the complete 10,000-image test set (Fig.~\ref{fig6}h). The input mask and offline-readout initialization are regenerated between realizations, while the test set remains fixed.

To isolate the contribution of conductance evolution, we repeat each realization with the memristive dynamics turned off (OFF-120) while keeping the input mask, dataset split and offline-readout settings unchanged. When the memristor states are held fixed, the mean accuracy decreases from $93.11\%$  to $11.43\%$, close to the $10\%$ chance level. This reduction shows that the time-dependent conductance evolution transforms the column sequence into discriminative final-state features. For comparison, reducing the 324 pixel values to 120 dimensions using principal-component analysis (PCA-120) or a fixed random projection (RP-120) gives mean accuracies of $92.14\%$ and $91.31\%$, respectively. Direct classification using all 324 resized pixels (RAW-324) gives $92.35\%$. The numerical reservoir therefore produces a compact 120-dimensional representation with performance comparable to that of the full pixel input.

The mean accuracy generally increases with the number of simulated units (Fig.~\ref{fig6}i), consistent with the availability of more independently transformed features. This model-level trend illustrates dimensional expansion by replicating the elementary-unit dynamics. Its translation to hardware additionally depends on device variability, interconnects and peripheral circuitry.

The input timing parameters are optimized within the numerical model. In the selected regime, $T_{\rm col}=2\,\mu\mathrm{s}$, and the 18-column sequence samples a restricted phase window of the $1\,\mathrm{kHz}$ carrier, allowing the conductance to evolve continuously across successive columns. This result identifies the relative timing between the column input and device-state evolution as a key design parameter. Its translation to hardware will require joint optimization of the drive waveform, device kinetics and peripheral-circuit bandwidth.

\section{Discussion}

This work establishes a modular memristive-circuit architecture in which a common elementary unit supports distinct system-level functions through programmable interaction, encoding, and readout rules. Its modular character arises from separating reusable state-bearing units from the circuit rules that organize their states. Experiments demonstrate electrically writable pseudospins, feedback-selected configurations, and symmetry-programmable transport. Circuit-model simulations examine domain-wall control and pattern recovery, while device-calibrated reservoir simulations demonstrate sequence-dependent feature generation. These results demonstrate how a reusable memristive unit supports distinct functions through programmable feedback, encoding, and readout.

A central physical consequence is that the pseudospin configuration alone does not determine the transport symmetry. The AFM- and AM-type configurations use the same measured compensated staggered pattern but assign different spatial relations to the sublattice hopping environments, producing distinct momentum- and real-space responses. Programmable feedback controls the formation and recovery of pseudospin configurations. The encoder then combines the measured conductance pattern with sublattice labels and the selected transport mode to configure the transport network, linking collective ordering to programmable transport.

In the continuous operating regime, the device-calibrated memristive model transforms sequentially encoded images into history-dependent final-state features. The loss of classification accuracy when conductance evolution is turned off identifies the time-dependent state update as the source of this transformation. Comparisons with principal-component and random-projection representations further indicate compact encoding of the complete input sequence. Reservoir-based processing therefore represents an additional operating mode of the elementary-unit dynamics. Its hardware realization will require matching the input timing and waveform to the device kinetics and peripheral-circuit bandwidth. More broadly, enriching the programmable state and coupling spaces could enable multilevel collective dynamics and nonreciprocal interactions. Combined with symmetry-selective transport encoding, these extensions may provide circuit platforms for exploring circuit analogs of altermagnetic topological models~\cite{Ma2024AltermagneticTI,Zhao2025WeylAltermagnet}. The modular architecture thus provides a common framework for state programming, collective dynamics, transport control, and information processing.

\section{Methods}

\subsection{Linear analysis of the feedback-coupled chain}

In the feedback-dominated stage, the external input and nonlinear term are neglected. For a periodic 1D chain with a uniform feedback sign $\xi$, the linearized dynamics are
\begin{equation}
\frac{dx_i}{d\tau}
=
-\mu x_i
+\lambda\xi\left(x_{i-1}+x_{i+1}\right),
\qquad
x_{i+N}=x_i.
\label{eq:linearized_periodic_chain}
\end{equation}
Using the Fourier expansion
\begin{equation}
x_i(\tau)
=
\frac{1}{\sqrt{N}}
\sum_q
\widetilde{x}_q(\tau)e^{\mathrm{i}qi},
\qquad
q=\frac{2\pi n}{N}.
\label{eq:chain_fourier_expansion}
\end{equation}
Eq.~\eqref{eq:linearized_periodic_chain} becomes
\begin{equation}
\frac{d\widetilde{x}_q}{d\tau}
=
\left(-\mu+2\lambda\xi\cos q\right)
\widetilde{x}_q.
\label{eq:chain_mode_equation}
\end{equation}
The corresponding mode growth rate is therefore Eq.~\eqref{eq:chain_growth_rate}.

\subsection{Initialization and statistical analysis of competing feedback}
To avoid favoring either collective mode initially, the voltage state was initialized as
\begin{equation}
V_i^{(0)}
=
A_U+A_S(-1)^{r_i+c_i}+\delta_i,
\label{eq:competition_initialization}
\end{equation}
where $A_U$ and $A_S$ were independently sampled zero-mean variables and $\delta_i$ was a weak site-dependent perturbation. For each value of $R_U$, 400 independently initialized realizations were evolved to convergence.

The crossover resistance $R_U^*$ was obtained from the equal-selection condition for the uniform and staggered states. Its uncertainty was estimated by bootstrap resampling. For the recovery analysis, $k$ sites were randomly flipped in either a complete uniform or complete staggered target pattern before the circuit evolution. For a target configuration $m_i^{\mathrm{tar}}$, the final overlap was defined as
\begin{equation}
Q=\left|\frac{1}{N}\sum_{i=1}^{N}m_i m_i^{\mathrm{tar}}\right|.
\end{equation}
A trajectory was counted as exactly recovered when $Q=1$. For each resistance, target pattern and error count, 400 independently initialized trajectories were simulated. The shared recovery reliability was defined as the smaller of the two 95\% Wilson lower confidence bounds. The region shown in Fig.~\ref{fig5}d contains the conditions under which this shared lower bound exceeds 0.90.

\subsection{Tight-binding calculations and symmetry analysis}

The three Hamiltonians obtained from the unified model are
\begin{align}
H_{\mathrm{FM}}(\mathbf{k})
&=
H_0(\mathbf{k})
+d(\mathbf{k})\sigma_0\tau_0
-\Delta\sigma_z\tau_0
-\delta\sigma_z\tau_z,
\label{eq:H_FM}\\
H_{\mathrm{AFM}}(\mathbf{k})
&=
H_0(\mathbf{k})
+d(\mathbf{k})\sigma_0\tau_0
+t_\perp\sigma_x\tau_0
-M\sigma_z\tau_z,
\label{eq:H_AFM}\\
H_{\mathrm{AM}}(\mathbf{k})
&=
H_0(\mathbf{k})
+d(\mathbf{k})\sigma_0\tau_z
+t_\perp\sigma_x\tau_0
-M\sigma_z\tau_z,
\label{eq:H_AM}
\end{align}
where $M=\Delta+\delta$.

For the AFM model, the bond-centered inversion and pseudospin reversal are represented internally by $\mathcal P=\tau_x$ and
$\mathcal T_m=\sigma_x\mathcal K$, respectively, where $\mathcal K$ denotes complex conjugation. Their product defines an effective antiunitary operation that leaves the Bloch momentum invariant:
\begin{equation}
\Theta_{\mathrm{AFM}}
=
\mathcal P\mathcal T_m
=
\sigma_x\tau_x\mathcal K,
\qquad
\Theta_{\mathrm{AFM}}^2=+1.
\label{eq:Theta_AFM_methods}
\end{equation}
It satisfies
\begin{equation}
\Theta_{\mathrm{AFM}}
H_{\mathrm{AFM}}(\mathbf{k})
\Theta_{\mathrm{AFM}}^{-1}
=
H_{\mathrm{AFM}}(\mathbf{k}).
\label{eq:AFM_same_momentum_symmetry}
\end{equation}
Thus, $\Theta_{\mathrm{AFM}}$ relates the two compensated sectors at the same momentum but does not by itself enforce their degeneracy.

In the decoupled limit $t_\perp=0$, the Hamiltonian acquires the additional unitary symmetry
\begin{equation}
[\sigma_z,H_{\mathrm{AFM}}(\mathbf{k})]=0.
\label{eq:AFM_sigmaz_conservation}
\end{equation}
Combining this conserved operator with $\Theta_{\mathrm{AFM}}$ gives
\begin{equation}
\Theta_{\mathrm{AFM}}^{(0)}
=
\sigma_z\Theta_{\mathrm{AFM}}
=
i\sigma_y\tau_x\mathcal K,
\qquad
\left[\Theta_{\mathrm{AFM}}^{(0)}\right]^2=-1.
\label{eq:AFM_decoupled_operator}
\end{equation}
This momentum-preserving antiunitary operation satisfies
\begin{equation}
\Theta_{\mathrm{AFM}}^{(0)}
H_{\mathrm{AFM}}(\mathbf{k})
\left[\Theta_{\mathrm{AFM}}^{(0)}\right]^{-1}
=
H_{\mathrm{AFM}}(\mathbf{k}),
\qquad
t_\perp=0,
\label{eq:AFM_decoupled_symmetry}
\end{equation}
and therefore enforces exact twofold degeneracy throughout the BZ. A finite $t_\perp$ breaks the conservation of $\sigma_z$ and removes this global degeneracy protection.

Writing $\epsilon_{\mathrm{AFM}}(\mathbf{k})
=\epsilon(\mathbf{k})+d(\mathbf{k})$,
the exact AFM eigenvalues are
\begin{equation}
E_{\alpha,\beta}(\mathbf{k})
=
\epsilon_{\mathrm{AFM}}(\mathbf{k})
+
\alpha
\sqrt{
M^2+
\left[
\gamma(\mathbf{k})+\beta t_\perp
\right]^2
},
\qquad
\alpha,\beta=\pm1.
\label{eq:AFM_finite_tperp_bands}
\end{equation}
For finite $t_\perp$, the splitting between the two branches with the same $\alpha$ is, to leading order,
\begin{equation}
\Delta E_{\mathrm{AFM}}(\mathbf{k})
\simeq
\frac{2|\gamma(\mathbf{k})t_\perp|}{M},
\qquad
|t_\perp|,|\gamma|\ll M.
\label{eq:AFM_weak_splitting}
\end{equation}

Along the boundary segments $X$--$M$ and $M$--$Y$,
\begin{equation}
\cos k_x+\cos k_y=0,
\qquad
\gamma(\mathbf{k})=0.
\label{eq:AFM_boundary_condition}
\end{equation}
The NN intersublattice coupling therefore vanishes, yielding the additional unitary symmetry
\begin{equation}
[\tau_z,H_{\mathrm{AFM}}(\mathbf{k})]=0,
\qquad
\mathbf{k}\in X\text{--}M\text{--}Y.
\label{eq:AFM_boundary_tauz}
\end{equation}
Combining this boundary symmetry with $\Theta_{\mathrm{AFM}}$ gives
\begin{equation}
\widetilde{\Theta}_{\mathrm{AFM}}
=
\tau_z\Theta_{\mathrm{AFM}}
=
i\sigma_x\tau_y\mathcal K,
\qquad
\widetilde{\Theta}_{\mathrm{AFM}}^{\,2}=-1.
\label{eq:AFM_boundary_operator_methods}
\end{equation}
On $X$--$M$--$Y$, it satisfies
\begin{equation}
\widetilde{\Theta}_{\mathrm{AFM}}
H_{\mathrm{AFM}}(\mathbf{k})
\widetilde{\Theta}_{\mathrm{AFM}}^{-1}
=
H_{\mathrm{AFM}}(\mathbf{k}).
\label{eq:AFM_boundary_symmetry_methods}
\end{equation}
This square-minus-one antiunitary symmetry protects the exact twofold degeneracy along $X$--$M$--$Y$ even for finite $t_\perp$. Away from this boundary, $\gamma(\mathbf{k})$ becomes finite, the
additional $\tau_z$ symmetry is lost, and the branches are weakly split according to Eq.~\eqref{eq:AFM_weak_splitting}.

For the AM model, the internal antiunitary transformation has the same matrix representation as in the AFM case,
\begin{equation}
\Theta_{\mathrm{AM}}
=
\sigma_x\tau_x\mathcal K.
\label{eq:Theta_AM_methods}
\end{equation}
Its symmetry action is nevertheless different because the sublattice-dependent term $d(\mathbf{k})\sigma_0\tau_z$ changes sign
under the sublattice exchange. The corresponding sign reversal is supplied by a fourfold rotation of the Bloch momentum,
\begin{equation}
C_{4z}(k_x,k_y)=(-k_y,k_x),
\label{eq:C4_momentum}
\end{equation}
under which
\begin{equation}
\begin{aligned}
\epsilon(C_{4z}\mathbf{k})&=\epsilon(\mathbf{k}),\\
\gamma(C_{4z}\mathbf{k})&=\gamma(\mathbf{k}),\\
d(C_{4z}\mathbf{k})&=-d(\mathbf{k}).
\end{aligned}
\label{eq:C4_transformation}
\end{equation}
Because $\tau_x\tau_z\tau_x=-\tau_z$, the AM term transforms as
\begin{equation}
\Theta_{\mathrm{AM}}
\left[
d(\mathbf{k})\sigma_0\tau_z
\right]
\Theta_{\mathrm{AM}}^{-1}
=
-d(\mathbf{k})\sigma_0\tau_z
=
d(C_{4z}\mathbf{k})\sigma_0\tau_z.
\label{eq:AM_term_transformation}
\end{equation}
The complete AM symmetry relation is therefore
\begin{equation}
\Theta_{\mathrm{AM}}
H_{\mathrm{AM}}(\mathbf{k})
\Theta_{\mathrm{AM}}^{-1}
=
H_{\mathrm{AM}}(C_{4z}\mathbf{k}).
\label{eq:AM_symmetry_methods}
\end{equation}
Unlike the AFM antiunitary symmetries above, this relation connects different momenta and therefore does not enforce same-momentum degeneracy. Instead, it permits pseudospin splitting at generic $\mathbf{k}$ while requiring
\begin{equation}
E_{+}(\mathbf{k})
=
E_{-}(C_{4z}\mathbf{k}),
\label{eq:AM_rotated_pairing}
\end{equation}
which produces the characteristic rotation-related AM spectral structure.

\subsection{Experimental measurements}
The memristive circuit unit was driven through its external input port using a 1-kHz $10~\mathrm{V_{pp}}$ sinusoidal voltage, generated by a Tektronix AFG31102 arbitrary-function generator. The analog circuit was powered by a SIGLENT SPD3303C dc power supply. The input voltage and the outputs of the individual functional stages were recorded using a Tektronix 3 Series Mixed Domain Oscilloscope. The memristive response was measured at the output of the AD844-based voltage-to-current conversion stage. The output voltage was converted into current according to the voltage-to-current conversion ratio and plotted against the simultaneously measured input voltage to obtain the current–voltage hysteresis loop. For coupled-unit measurements, the internal output of one unit was connected to the internal input of a neighboring unit through either the direct or inverter-assisted feedback path. The state-dependent outputs of the coupled units were recorded to identify uniform and staggered pseudospin configurations.

The binary-state maps used for transport programming were obtained from the memristive-state measurements described above. After the memristive-state measurements were completed, the resulting $6\times 6$ binary-state maps were transferred offline to the MCU control program. The MCU then combined $m_i$ with prescribed sublattice label $\eta_i$ and the selected transport rule $\chi$ to determine $Q_i^\chi$ and $P_i^\chi$ at each site. The resulting selector values configured the diagonal-coupling and grounding branches of the $6\times6$ transport layer. The lattice geometry and NN connections remained unchanged among the three configurations. For local excitation, a 3-V dc source supplied by a SIGLENT SPD3303C power supply was connected to the selected lattice node through a $1\,\mathrm{k\Omega}$ series resistor, while the same node was connected to ground through an additional $1\,\mathrm{k\Omega}$ resistor. After the circuit reached a steady state, the voltages at all 36 nodes were measured relative to the common ground using a Tektronix 3 Series Mixed Domain Oscilloscope. Excitation at representative neighboring $A$- and $B$-sublattice sites yielded the real-space voltage distributions shown in Fig.~\ref{fig4}. To reconstruct the complete circuit response, each of the 36 nodes was excited sequentially, and the resulting 36-node voltage distribution was recorded for each excitation position. The programmed lattice configuration was held fixed throughout each measurement set.

\subsection{Experimental reconstruction of the circuit-Laplacian spectra}

The transport experiments were performed on a $6\times6$ resistor lattice comprising two sublattices, $A$ and $B$. The diagonal and grounding branches were programmed according to the selected FM-, AFM- or AM-type encoding rule, while the lattice geometry and nearest-neighbor network were kept unchanged. For a programmed configuration $\chi$, the steady-state circuit response is described by
\begin{equation}
\mathbf I=\mathbf J_{\chi}\mathbf V,
\label{eq:circuit_response}
\end{equation}
where $\mathbf V$ and $\mathbf I$ are the node-voltage and externally injected-current vectors, respectively, and $\mathbf J_{\chi}$ is the finite-circuit Laplacian.

For the real-space voltage measurements, a selected node was connected to a $3~\mathrm{V}$ source through a series resistor $R_{\mathrm{s}}=1~\mathrm{k}\Omega$ and to ground through a grounding resistor $R_{\mathrm{g}}=1~\mathrm{k}\Omega$. The steady-state voltages at all 36 lattice nodes were then recorded relative to ground. Measurements with the source applied to representative $A$- and $B$-sublattice nodes yielded the
real-space voltage profiles shown in the main text.

To reconstruct the complete experimental response, the same measurement was repeated with each of the 36 nodes serving sequentially as the source node. For excitation at node $i$, the net current injected into the transport lattice was determined from the measured source-node voltage as
\begin{equation}
I_i=
\frac{V_{\mathrm{s}}-V_i^{(i)}}{R_{\mathrm{s}}}
-\frac{V_i^{(i)}}{R_{\mathrm{g}}},
\qquad
V_{\mathrm{s}}=3~\mathrm{V},
\qquad
R_{\mathrm{s}}=R_{\mathrm{g}}=1~\mathrm{k}\Omega,
\label{eq:injected_current}
\end{equation}
where $V_i^{(i)}$ denotes the voltage measured at node $i$ when that node is excited. The first term is the current supplied through $R_{\mathrm{s}}$, whereas the second is the current diverted to ground through $R_{\mathrm{g}}$. The remaining current is therefore injected into the transport lattice. The measured voltages define the corresponding column of the impedance matrix according to
\begin{equation}
Z_{ji}=\frac{V_j^{(i)}}{I_i},
\qquad i,j=1,\ldots,36,
\label{eq:impedance_element}
\end{equation}
where $V_j^{(i)}$ is the voltage measured at node $j$ under excitation at node $i$. Sequential excitation of all source nodes yielded the complete $36\times36$ experimental impedance matrix $\mathbf Z$.
The finite-circuit Laplacian $\mathbf J_{\mathrm{rec}}$ was reconstructed by computing the truncated singular-value-decomposition inverse of $\mathbf Z$. Singular values smaller than $10^{-10}$ times the largest singular value were discarded to suppress numerically unstable contributions to the inversion. A cutoff-dependence analysis confirmed that the value $10^{-10}$ lies within the full-rank interval and does not discard any measured singular mode (Supplemental Materials~\cite{suppmat}).

To construct an effective bulk description, each node was assigned a sublattice label $\alpha\in\{A, B\}$ and a lattice position $\mathbf r_i$. Matrix elements associated with the same ordered pair of sublattices and the
same relative displacement $\Delta\mathbf r=\mathbf r_j-\mathbf r_i$ were spatially averaged:
\begin{equation}
\overline{J}_{\alpha\beta}(\Delta\mathbf r)
=
\frac{1}{N_{\alpha\beta}(\Delta\mathbf r)}
\sum_{\substack{
i\in\alpha,\;j\in\beta\\
\mathbf r_j-\mathbf r_i=\Delta\mathbf r
}}
\left(J_{\mathrm{rec}}\right)_{ij},
\label{eq:spatial_average}
\end{equation}
where $N_{\alpha\beta}(\Delta\mathbf r)$ is the number of matrix elements in the corresponding displacement class. The effective two-sublattice Bloch circuit Laplacian was then constructed as
\begin{equation}
J_{\alpha\beta}(\mathbf k)
=
\sum_{\Delta\mathbf r}
\overline{J}_{\alpha\beta}(\Delta\mathbf r)
e^{\mathrm i\mathbf k\cdot\Delta\mathbf r},
\qquad
\alpha,\beta\in\{A,B\},
\label{eq:bloch_laplacian}
\end{equation}

A common two-node electrical unit cell with primitive vectors
\begin{equation}
\mathbf a_1=(1,1),
\qquad
\mathbf a_2=(1,-1)
\label{eq:primitive_vectors}
\end{equation}
were used for the FM-, AFM-, and AM-type configurations, allowing their experimentally reconstructed spectra to be compared within the same folded Brillouin zone. The spectral branches were obtained by solving
\begin{equation}
\mathbf J(\mathbf k)\boldsymbol{\phi}_n(\mathbf k)
=
\lambda_n(\mathbf k)\boldsymbol{\phi}_n(\mathbf k)
\label{eq:bloch_eigenproblem}
\end{equation}
along the path $\Gamma$--$X$--$M$--$Y$--$\Gamma$, with
\begin{equation}
\Gamma=(0,0),\quad
X=(-\pi/2,\pi/2),\quad
M=(0,\pi),\quad
Y=(\pi/2,\pi/2).
\label{eq:high_symmetry_points}
\end{equation}
The resulting eigenvalues have units of conductance and represent the circuit-Laplacian spectral branches of the experimentally programmed transport circuits.

\subsection{Reservoir-computing protocol}

The MNIST images were resized from $28\times28$ to $18\times18$ pixels using bilinear interpolation, with the pixel intensities retained in the range $[0,1]$. From the official training set, 48,000 images were used for fitting and 8,000 disjoint images for validation. The official 10,000-image test set was reserved for final evaluation. Each resized image was presented column by column as a sequence of 18 vectors, 
\begin{equation} 
\mathbf{p}(k)\in\mathbb{R}^{18}, \qquad k=1,\ldots,18. 
\end{equation}
At each input step, the image column was projected onto $N_{\mathrm{mem}}$ independently operated memristive units through a sparse random mask, 
\begin{equation} 
\mathbf{B}_{\mathrm{in}} \in\mathbb{R}^{N_{\mathrm{mem}}\times18}. \end{equation}
Each row of $\mathbf{B}_{\mathrm{in}}$ contained four randomly selected nonzero elements with independently assigned signs, \begin{equation} 
B_{ij}\in \left\{-\frac{1}{4},\,0,\,+\frac{1}{4}\right\}. \end{equation}
The masked input received by the $i$-th unit at step $k$ was \begin{equation} 
s_i(k)=\sum_{j=1}^{18}B_{ij}p_j(k), 
\end{equation} 
which modulated the amplitude of the applied sinusoidal voltage, \begin{equation} 
A_i(k)=V_0+\alpha s_i(k). 
\end{equation}
The corresponding source voltage was 
\begin{equation} 
V_{\mathrm{src},i}(t) = A_i(k) \sin\left[2\pi f\left(t+t_{\mathrm{start}}\right)\right], \qquad t\in[(k-1)T_{\mathrm{col}},kT_{\mathrm{col}}),
\end{equation} 
where $t$ denotes the elapsed time from the beginning of the image sequence. The carrier amplitude and frequency were fixed at $V_0=5.0~\mathrm{V}$ and $f=1~\mathrm{kHz}$, respectively. The input gain, column duration, and carrier phase were selected using the validation set. The resulting protocol used $\alpha=0.5~\mathrm{V}$, $T_{\mathrm{col}}=2~\mu\mathrm{s}$ and a normalized carrier phase of $\phi_{\mathrm{c}}=0.18$ at the center of the complete input window. This gives $t_{\mathrm{start}}=162~\mu\mathrm{s}$.

Before each image, all units were initialized to the same normalized conductance state, $g_i(0)=1$. The state evolved continuously across the 18 image columns, with no reset between columns and no state carried over between successive images. At the end of the final column, the current of each unit was sampled directly and assembled into the reservoir feature vector,
\begin{equation}
\mathbf{x}_{\mathrm{res}}
=
\left[
I_{1,\mathrm{end}},
I_{2,\mathrm{end}},
\ldots,
I_{N_{\mathrm{mem}},\mathrm{end}}
\right]^{\mathrm T}.
\end{equation}
The 120-unit reservoir therefore produced one 120-dimensional end-of-sequence feature vector for each image.

To quantify the class separability of the extracted reservoir features, classification was performed offline in software using a linear output layer that mapped the 120 inputs onto ten class logits, followed by a softmax operation. Each feature was standardized using the mean and standard deviation obtained from the fitting set. The parameters of the linear readout were determined using the fitting data, with hyperparameter selection performed by five-fold cross-validation. The input mask and memristive dynamics remained fixed throughout this offline analysis, and no label-dependent parameter updates were applied to the memristive model.

Four reference representations were evaluated using the same dataset partition and linear readout. In the state-evolution-off control (OFF-120), the masked-input procedure was retained while the conductance state was held fixed. For PCA-120, each flattened 324-dimensional image was projected onto the 120 leading principal components determined from the fitting data. For RP-120, the flattened image was projected using
\begin{equation}
\mathbf{x}_{\mathrm{RP}}
=
\mathbf{W}_{\mathrm{RP}}
\mathbf{x}_{\mathrm{raw}},
\qquad
\mathbf{W}_{\mathrm{RP}}
\in\mathbb{R}^{120\times324},
\end{equation}
where
\begin{equation}
(W_{\mathrm{RP}})_{ij}
\sim\mathcal{N}\left(0,\frac{1}{120}\right).
\end{equation}
Each random-projection matrix was generated once and applied to the fitting, validation, and test sets. For RAW-324, all 324 resized pixel values were supplied directly to the classifier. All representations used the same linear softmax readout and hyperparameter-selection procedure. All five representations were evaluated over ten independently seeded runs using the same fixed training, validation, and test partitions. Each run was evaluated on the complete official MNIST test set of 10,000 images. For ON-120 and OFF-120, the runs were paired: each pair used the same random input mask and classifier-training seed, and differed only in whether memristive state evolution was enabled. For PCA-120 and RAW-324, the input representations remained fixed across runs, and only the classifier-training seed was varied. For RP-120, both the random-projection matrix and the classifier-training seed were regenerated between runs. The accuracies reported in Fig.~\ref{fig6}h are the means over these ten runs.

For the reservoir-size analysis, $N_{\mathrm{mem}}$ was varied while the input dimension and mask sparsity were kept unchanged. Each reservoir size was evaluated over ten independent numerical realizations with different random input masks and classifier-training seeds. Within each realization, reservoirs of different sizes were constructed as nested subsets of the same maximum-size reservoir, enabling the size dependence to be compared under matched random conditions. Each point in Fig.~\ref{fig6}i represents the mean test accuracy over the ten realizations, and the shaded band denotes one standard deviation across realizations.

\section{Acknowledgements}

S.Q., Y.W., Z.H., S.L. and T.J.C. acknowledge support from the National Key Research and Development Program of China (Grant No. 2022YFA1404903) and the National Natural Science Foundation of China (Grant Nos. 62288101 and U22A2001). X.G. acknowledges support from the Young Scientists Fund of the National Natural Science Foundation of China (Grant No. 6260010130). Y.J. and C.S. acknowledge support from the Chinese Academy of Sciences (Project Nos. E4BA270100, E4Z127010F, E4Z6270100 and E53327020D). R.T. was funded by the Deutsche Forschungsgemeinschaft (DFG, German Research Foundation) – Project-ID 258499086 – SFB 1170; through the Würzburg-Dresden Cluster of Excellence on Complexity, Topology, and Dynamics in Quantum Materials (ctd.qmat) – Project-ID 390858490 – EXC 2147; and through the Research Unit QUAST – Project-ID 449872909 – FOR 5249.

\section{Data availability}

The datasets generated and analyzed in this study are available from the corresponding author upon reasonable request.

\bibliography{references}

@misc{suppmat,
  note = {Supplementary Materials}
}

@article{Strogatz2001ComplexNetworks,
  author  = {Strogatz, Steven H.},
  title   = {{Exploring Complex Networks}},
  journal = {Nature},
  volume  = {410},
  number  = {6825},
  pages   = {268--276},
  year    = {2001},
  doi     = {10.1038/35065725}
}

@article{Mott1936,
  author  = {Mott, Nevill Francis},
  title   = {{The Electrical Conductivity of Transition Metals}},
  journal = {Proc. R. Soc. Lond. A},
  volume  = {153},
  number  = {880},
  pages   = {699--717},
  year    = {1936},
  doi     = {10.1098/rspa.1936.0031}
}

@article{Fert1976,
  author  = {Fert, A. and Campbell, I. A.},
  title   = {{Electrical Resistivity of Ferromagnetic Nickel and Iron Based Alloys}},
  journal = {J. Phys. F: Met. Phys.},
  volume  = {6},
  number  = {5},
  pages   = {849--871},
  year    = {1976},
  doi     = {10.1088/0305-4608/6/5/025}
}

@article{Tedrow1973,
  author  = {Tedrow, P. M. and Meservey, R.},
  title   = {{Spin Polarization of Electrons Tunneling from Films of Fe, Co, Ni, and Gd}},
  journal = {Phys. Rev. B},
  volume  = {7},
  number  = {1},
  pages   = {318--326},
  year    = {1973},
  doi     = {10.1103/PhysRevB.7.318}
}

@article{Baibich1988,
  author  = {Baibich, M. N. and Broto, J. M. and Fert, A. and Van Dau, F. Nguyen and Petroff, F. and Etienne, P. and Creuzet, G. and Friederich, A. and Chazelas, J.},
  title   = {{Giant Magnetoresistance of (001)Fe/(001)Cr Magnetic Superlattices}},
  journal = {Phys. Rev. Lett.},
  volume  = {61},
  number  = {21},
  pages   = {2472--2475},
  year    = {1988},
  doi     = {10.1103/PhysRevLett.61.2472}
}

@article{Fert2008,
  author  = {Fert, Albert},
  title   = {{Nobel Lecture: Origin, Development, and Future of Spintronics}},
  journal = {Rev. Mod. Phys.},
  volume  = {80},
  number  = {4},
  pages   = {1517--1530},
  year    = {2008},
  doi     = {10.1103/RevModPhys.80.1517}
}

@article{Baltz2018,
  author  = {Baltz, V. and Manchon, A. and Tsoi, M. and Moriyama, T. and Ono, T. and Tserkovnyak, Y.},
  title   = {{Antiferromagnetic Spintronics}},
  journal = {Rev. Mod. Phys.},
  volume  = {90},
  number  = {1},
  pages   = {015005},
  year    = {2018},
  doi     = {10.1103/RevModPhys.90.015005}
}

@article{Ramazashvili2008,
  author  = {Ramazashvili, Revaz},
  title   = {{Kramers Degeneracy in a Magnetic Field and Zeeman Spin-Orbit Coupling in Antiferromagnetic Conductors}},
  journal = {Phys. Rev. Lett.},
  volume  = {101},
  number  = {13},
  pages   = {137202},
  year    = {2008},
  doi     = {10.1103/PhysRevLett.101.137202}
}

@article{Smejkal2018,
  author  = {{\v{S}}mejkal, Libor and Mokrousov, Yuriy and Yan, Binghai and MacDonald, Allan H.},
  title   = {{Topological Antiferromagnetic Spintronics}},
  journal = {Nat. Phys.},
  volume  = {14},
  number  = {3},
  pages   = {242--251},
  year    = {2018},
  doi     = {10.1038/s41567-018-0064-5}
}

@article{Smejkal2022,
  author  = {{\v{S}}mejkal, Libor and Sinova, Jairo and Jungwirth, Tom{\'a}{\v{s}}},
  title   = {{Beyond Conventional Ferromagnetism and Antiferromagnetism: A Phase with Nonrelativistic Spin and Crystal Rotation Symmetry}},
  journal = {Phys. Rev. X},
  volume  = {12},
  number  = {3},
  pages   = {031042},
  year    = {2022},
  doi     = {10.1103/PhysRevX.12.031042}
}

@article{Sinova2022,
  author  = {{\v{S}}mejkal, Libor and Sinova, Jairo and Jungwirth, Tom{\'a}{\v{s}}},
  title   = {{Emerging Research Landscape of Altermagnetism}},
  journal = {Phys. Rev. X},
  volume  = {12},
  number  = {4},
  pages   = {040501},
  year    = {2022},
  doi     = {10.1103/PhysRevX.12.040501}
}

@article{Jungwirth2026,
  author  = {Jungwirth, Tom{\'a}{\v{s}} and Sinova, Jairo and Fernandes, Rafael M. and Liu, Qihang and Watanabe, Hikaru and Murakami, Shuichi and Nakatsuji, Satoru and {\v{S}}mejkal, Libor},
  title   = {{Symmetry, Microscopy and Spectroscopy Signatures of Altermagnetism}},
  journal = {Nature},
  volume  = {649},
  number  = {8098},
  pages   = {837--847},
  year    = {2026},
  doi     = {10.1038/s41586-025-09883-2}
}

@article{Song2025,
  author  = {Song, Cheng and Bai, Hua and Zhou, Zhiyuan and Han, Lei and Reichlova, Helena and Dil, J. Hugo and Liu, Junwei and Chen, Xianzhe and Pan, Feng},
  title   = {{Altermagnets as a new class of functional materials}},
  journal = {Nat. Rev. Mater.},
  volume  = {10},
  number  = {6},
  pages   = {473--485},
  year    = {2025},
  doi     = {10.1038/s41578-025-00779-1}
}

@article{Zhou2024,
  author  = {Zhou, Xiaodong and Feng, Wanxiang and Zhang, Run-Wu and \ifmmode \check{S}\else \v{S}\fi{}mejkal, Libor and Sinova, Jairo and Mokrousov, Yuriy and Yao, Yugui},
  title   = {{Crystal Thermal Transport in Altermagnetic ${\mathrm{RuO}}_{2}$}},
  journal = {Phys. Rev. Lett.},
  volume  = {132},
  number  = {5},
  pages   = {056701},
  year    = {2024},
  doi     = {10.1103/PhysRevLett.132.056701}
}

@article{Krempasky2024,
  author  = {Krempaský, J. and Šmejkal, L. and D’Souza, S. W. and Hajlaoui, M. and Springholz, G. and Uhlířová, K. and Alarab, F. and Constantinou, P. C. and Strocov, V. and Usanov, D. and Pudelko, W. R. and González-Hernández, R. and Birk Hellenes, A. and Jansa, Z. and Reichlová, H. and Šobáň, Z. and Gonzalez Betancourt, R. D. and Wadley, P. and Sinova, J. and Kriegner, D. and Minár, J. and Dil, J. H. and Jungwirth, T.},
  title   = {{Altermagnetic lifting of Kramers spin degeneracy}},
  journal = {Nature},
  volume  = {626},
  number  = {7999},
  pages   = {517--522},
  year    = {2024},
  doi     = {10.1038/s41586-023-06907-7}
}

@article{Amin2024,
  author  = {Amin, O. J. and Dal Din, A. and Golias, E. and Niu, Y. and Zakharov, A. and Fromage, S. C. and Fields, C. J. B. and Heywood, S. L. and Cousins, R. B. and Maccherozzi, F. and Krempaský, J. and Dil, J. H. and Kriegner, D. and Kiraly, B. and Campion, R. P. and Rushforth, A. W. and Edmonds, K. W. and Dhesi, S. S. and Šmejkal, L. and Jungwirth, T. and Wadley, P.},
  title   = {{Nanoscale imaging and control of altermagnetism in MnTe}},
  journal = {Nature},
  volume  = {636},
  number  = {8042},
  pages   = {348--353},
  year    = {2024},
  doi     = {10.1038/s41586-024-08234-x}
}

@article{Igor2004,
  author  = {\ifmmode \check{Z}\else \v{Z}\fi{}uti\ifmmode \acute{c}\else \'{c}\fi{}, Igor and Fabian, Jaroslav and Das Sarma, S.},
  title   = {{Spintronics: Fundamentals and applications}},
  journal = {Rev. Mod. Phys.},
  volume  = {76},
  number  = {2},
  pages   = {323--410},
  year    = {2004},
  doi     = {10.1103/RevModPhys.76.323}
}

@article{Jungwirth2016,
  author  = {Jungwirth, T. and Marti, X. and Wadley, P. and Wunderlich, J.},
  title   = {{Antiferromagnetic spintronics}},
  journal = {Nat. Nanotechnol.},
  volume  = {11},
  number  = {3},
  pages   = {231--241},
  year    = {2016},
  doi     = {10.1038/nnano.2016.18}
}

@article{Wadley2018,
  author  = {Železný, J. and Wadley, P. and Olejník, K. and Hoffmann, A. and Ohno, H.},
  title   = {{Spin transport and spin torque in antiferromagnetic devices}},
  journal = {Nat. Phys.},
  volume  = {14},
  number  = {3},
  pages   = {220--228},
  year    = {2018},
  doi     = {10.1038/s41567-018-0062-7}
}

@article{Jungwirth2018,
  author  = {Jungwirth, T. and Sinova, J. and Manchon, A. and Marti, X. and Wunderlich, J. and Felser, C.},
  title   = {{The multiple directions of antiferromagnetic spintronics}},
  journal = {Nat. Phys.},
  volume  = {14},
  number  = {3},
  pages   = {200--203},
  year    = {2018},
  doi     = {10.1038/s41567-018-0063-6}
}

@article{Helbig2019,
  author  = {Helbig, Tobias and Hofmann, Tobias and Lee, Ching Hua and Thomale, Ronny and Imhof, Stefan and Molenkamp, Laurens W. and Kiessling, Tobias},
  title   = {{Band structure engineering and reconstruction in electric circuit networks}},
  journal = {Phys. Rev. B},
  volume  = {99},
  number  = {16},
  pages   = {161114(R)},
  year    = {2019},
  doi     = {10.1103/PhysRevB.99.161114}
}

@article{Helbig2020,
  author  = {Helbig, T. and Hofmann, T. and Imhof, S. and Abdelghany, M. and Kiessling, T. and Molenkamp, L. W. and Lee, C. H. and Szameit, A. and Greiter, M. and Thomale, R.},
  title   = {{Generalized bulk–boundary correspondence in non-Hermitian topolectrical circuits}},
  journal = {Nat. Phys.},
  volume  = {16},
  number  = {7},
  pages   = {747--750},
  year    = {2020},
  doi     = {10.1038/s41567-020-0922-9}
}

@article{Qiu2025,
  author  = {Qiu, Sichang and Hu, Jinbing and Yang, Yi and Shang, Ce and Liu, Shuo and Cui, Tie Jun},
  title   = {{Octupole topological insulating phase protected by a three-dimensional momentum-space nonsymmorphic group}},
  journal = {Natl. Sci. Rev.},
  volume  = {12},
  number  = {7},
  pages   = {nwaf137},
  year    = {2025},
  doi     = {10.1093/nsr/nwaf137}
}

@article{Jia2026,
 author = {Jia, Hao and Yang, Shanglin and He, Jiajun and Liu, Shuo and Chen, Haoxiang and Shang, Ce and Ma, Shaojie and Han, Peng and Lee, Ching Hua and Gao, Zhen and Lai, Yun and Cui, Tie Jun},
   title = {Deep-learning-empowered programmable topolectrical circuits},
   journal = {Nature Communications},
   volume = {17},
   number = {1},
   pages = {6144},
   ISSN = {2041-1723},
   DOI = {10.1038/s41467-026-72901-y},
   url = {https://doi.org/10.1038/s41467-026-72901-y},
   year = {2026},
   type = {Journal Article}
}

@article{Onsager1944,
  author  = {Onsager, Lars},
  title   = {{Crystal Statistics. I. A Two-Dimensional Model with an Order-Disorder Transition}},
  journal = {Phys. Rev.},
  volume  = {65},
  number  = {3-4},
  pages   = {117--149},
  year    = {1944},
  doi     = {10.1103/PhysRev.65.117}
}

@article{Glauber1963,
  author  = {Glauber, Roy J.},
  title   = {Time-Dependent Statistics of the Ising Model},
  journal = {J. Math. Phys.},
  volume  = {4},
  number  = {2},
  pages   = {294--307},
  year    = {1963},
  doi     = {10.1063/1.1703954}
}

@article{Pershin2009,
  author  = {Pershin, Yuriy V. and La Fontaine, Steven and Di Ventra, Massimiliano},
  title   = {{Memristive model of amoeba learning}},
  journal = {Phys. Rev. E},
  volume  = {80},
  number  = {2},
  pages   = {021926},
  year    = {2009},
  doi     = {10.1103/PhysRevE.80.021926}
}

@article{Dowling2022,
  author  = {Dowling, Vincent J. and Pershin, Yuriy V.},
  title   = {{Memristive Ising circuits}},
  journal = {Phys. Rev. E},
  volume  = {106},
  number  = {5},
  pages   = {054156},
  year    = {2022},
  doi     = {10.1103/PhysRevE.106.054156}
}

@article{Jiang2026,
   author = {Jiang, Yuxin and Li, Wenhao and Zhang, Weixuan and Shang, Ce and Yang, Yihao and Zhang, Xiangdong and Lee, Ching Hua and Thomale, Ronny},
   title = {Engineering topolectrical circuits},
   journal = {Nat. Rev. Electr. Eng.},
   ISSN = {2948-1201},
   DOI = {10.1038/s44287-026-00308-4},
   url = {https://doi.org/10.1038/s44287-026-00308-4},
   year = {2026},
   type = {Journal Article}
}

@article{Mazin2022,
  author  = {Mazin, Igor and {The PRX Editors}},
  title   = {{Editorial: Altermagnetism---A New Punch Line of Fundamental Magnetism}},
  journal = {Phys. Rev. X},
  volume  = {12},
  number  = {4},
  pages   = {040002},
  year    = {2022},
  doi     = {10.1103/physrevx.12.040002}
}

@article{GonzalezHernandez2021,
  author  = {González-Hernández, Rafael and Šmejkal, Libor and Výborný, Karel and Yahagi, Yuta and Sinova, Jairo and Jungwirth, Tomáš and Železný, Jakub},
  title   = {{Efficient Electrical Spin Splitter Based on Nonrelativistic Collinear Antiferromagnetism}},
  journal = {Phys. Rev. Lett.},
  volume  = {126},
  number  = {12},
  pages   = {127701},
  year    = {2021},
  doi     = {10.1103/physrevlett.126.127701}
}

@article{Smejkal2020,
  author  = {Šmejkal, Libor and González-Hernández, Rafael and Jungwirth, T. and Sinova, J.},
  title   = {{Crystal time-reversal symmetry breaking and spontaneous Hall effect in collinear antiferromagnets}},
  journal = {Sci. Adv.},
  volume  = {6},
  number  = {23},
  pages   = {eaaz8809},
  year    = {2020},
  doi     = {10.1126/sciadv.aaz8809}
}

@article{Feng2022,
  author  = {Feng, Zexin and Zhou, Xiaorong and Šmejkal, Libor and Wu, Lei and Zhu, Zengwei and Guo, Huixin and González-Hernández, Rafael and Wang, Xiaoning and Yan, Han and Qin, Peixin and Zhang, Xin and Wu, Haojiang and Chen, Hongyu and Meng, Ziang and Liu, Li and Xia, Zhengcai and Sinova, Jairo and Jungwirth, Tomáš and Liu, Zhiqi},
  title   = {{An anomalous Hall effect in altermagnetic ruthenium dioxide}},
  journal = {Nat. Electron.},
  volume  = {5},
  number  = {11},
  pages   = {735--743},
  year    = {2022},
  doi     = {10.1038/s41928-022-00866-z}
}

@article{McGuire1975,
  author  = {McGuire, T. and Potter, R.},
  title   = {{Anisotropic magnetoresistance in ferromagnetic {3d} alloys}},
  journal = {IEEE Trans. Magn.},
  volume  = {11},
  number  = {4},
  pages   = {1018--1038},
  year    = {1975},
  doi     = {10.1109/tmag.1975.1058782}
}

@article{Albert2015,
  author  = {Albert, Victor V. and Glazman, Leonid I. and Jiang, Liang},
  title   = {{Topological Properties of Linear Circuit Lattices}},
  journal = {Phys. Rev. Lett.},
  volume  = {114},
  number  = {17},
  pages   = {173902},
  year    = {2015},
  doi     = {10.1103/physrevlett.114.173902}
}

@article{Ningyuan2015,
  author  = {Ningyuan, Jia and Owens, Clai and Sommer, Ariel and Schuster, David and Simon, Jonathan},
  title   = {{Time- and Site-Resolved Dynamics in a Topological Circuit}},
  journal = {Phys. Rev. X},
  volume  = {5},
  number  = {2},
  pages   = {021031},
  year    = {2015},
  doi     = {10.1103/physrevx.5.021031}
}

@article{Imhof2018,
  author  = {Imhof, Stefan and Berger, Christian and Bayer, Florian and Brehm, Johannes and Molenkamp, Laurens W. and Kiessling, Tobias and Schindler, Frank and Lee, Ching Hua and Greiter, Martin and Neupert, Titus and Thomale, Ronny},
  title   = {{Topolectrical-circuit realization of topological corner modes}},
  journal = {Nat. Phys.},
  volume  = {14},
  number  = {9},
  pages   = {925--929},
  year    = {2018},
  doi     = {10.1038/s41567-018-0246-1}
}

@article{Lee2018,
  author  = {Lee, Ching Hua and Imhof, Stefan and Berger, Christian and Bayer, Florian and Brehm, Johannes and Molenkamp, Laurens W. and Kiessling, Tobias and Thomale, Ronny},
  title   = {{Topolectrical Circuits}},
  journal = {Commun. Phys.},
  volume  = {1},
  number  = {1},
  pages   = {39},
  year    = {2018},
  doi     = {10.1038/s42005-018-0035-2}
}

@article{Ezawa2018,
  author  = {Ezawa, Motohiko},
  title   = {{Higher-order topological electric circuits and topological corner resonance on the breathing kagome and pyrochlore lattices}},
  journal = {Phys. Rev. B},
  volume  = {98},
  number  = {20},
  pages   = {201402},
  year    = {2018},
  doi     = {10.1103/physrevb.98.201402}
}

@article{Yang2013,
  author  = {Yang, J. Joshua and Strukov, Dmitri B. and Stewart, Duncan R.},
  title   = {{Memristive devices for computing}},
  journal = {Nat. Nanotechnol.},
  volume  = {8},
  number  = {1},
  pages   = {13--24},
  year    = {2013},
  doi     = {10.1038/nnano.2012.240}
}

@article{Zidan2018,
  author  = {Zidan, Mohammed A. and Strachan, John Paul and Lu, Wei D.},
  title   = {{The future of electronics based on memristive systems}},
  journal = {Nat. Electron.},
  volume  = {1},
  number  = {1},
  pages   = {22--29},
  year    = {2018},
  doi     = {10.1038/s41928-017-0006-8}
}

@article{Ma2024AltermagneticTI,
  author  = {Ma, Hai-Yang and Jia, Jin-Feng},
  title   = {{Altermagnetic topological insulator and the selection rules}},
  journal = {Phys. Rev. B},
  volume  = {110},
  pages   = {064426},
  year    = {2024},
  doi     = {10.1103/PhysRevB.110.064426}
}

@article{Zhao2025WeylAltermagnet,
  author  = {Zhao, Yufei and Mao, Zhiqiang and Yan, Binghai},
  title   = {{Nonlinear transport signatures of hidden symmetry breaking in a Weyl altermagnet}},
  journal = {Phys. Rev. B},
  volume  = {112},
  pages   = {165127},
  year    = {2025},
  doi     = {10.1103/f3vr-x4fx}
}

@article{Rikvold2016,
  title = {Equilibrium, metastability, and hysteresis in a model spin-crossover material with nearest-neighbor antiferromagnetic-like and long-range ferromagnetic-like interactions},
  author = {Rikvold, Per Arne and Brown, Gregory and Miyashita, Seiji and Omand, Conor and Nishino, Masamichi},
  journal = {Phys. Rev. B},
  volume = {93},
  issue = {6},
  pages = {064109},
  numpages = {13},
  year = {2016},
  month = {Feb},
  publisher = {American Physical Society},
  doi = {10.1103/PhysRevB.93.064109},
  url = {https://link.aps.org/doi/10.1103/PhysRevB.93.064109}
}

@article{Du2017RC,
  author  = {Du, Chao and Cai, Fuxi and Zidan, Mohammed A. and Ma, Wen
             and Lee, Seung Hwan and Lu, Wei D.},
  title   = {Reservoir Computing Using Dynamic Memristors for Temporal
             Information Processing},
  journal = {Nat. Commun.},
  volume  = {8},
  pages   = {2204},
  year    = {2017},
  doi     = {10.1038/s41467-017-02337-y}
}

@article{Moon2019RC,
  author  = {Moon, John and Ma, Wen and Shin, Jong Hoon and Cai, Fuxi
             and Du, Chao and Lee, Seung Hwan and Lu, Wei D.},
  title   = {Temporal Data Classification and Forecasting Using a
             Memristor-Based Reservoir Computing System},
  journal = {Nat. Electron.},
  volume  = {2},
  pages   = {480--487},
  year    = {2019},
  doi     = {10.1038/s41928-019-0313-3}
}

@article{Tanaka2019PRC,
  author  = {Tanaka, G. and Yamane, T. and H{\'e}roux, J.-B. and Nakane, R. and Kanazawa, N. and Takeda, S. and Numata, H. and Nakano, D. and Hirose, A.},
  title   = {Recent advances in physical reservoir computing: A review},
  journal = {Neural Netw.},
  volume  = {115},
  pages   = {100--123},
  year    = {2019},
  doi     = {10.1016/j.neunet.2019.03.005}
}

@article{Zhong2021RC,
  author  = {Zhong, Yanan and Tang, Jianshi and Li, Xinyi and Gao, Bin
             and Qian, He and Wu, Huaqiang},
  title   = {Dynamic Memristor-Based Reservoir Computing for High-Efficiency Temporal Signal Processing},
  journal = {Nat. Commun.},
  volume  = {12},
  pages   = {408},
  year    = {2021},
  doi     = {10.1038/s41467-020-20692-1}
}

@article{Spagnolo2022PhotonicMemristor,
  author  = {Spagnolo, M. and Morris, J. and Piacentini, S. and Antesberger, M. and Massa, F. and Crespi, A. and Ceccarelli, F. and Osellame, R. and Walther, P.},
  title   = {Experimental photonic quantum memristor},
  journal = {Nat. Photonics},
  volume  = {16},
  pages   = {318--323},
  year    = {2022},
  doi     = {10.1038/s41566-022-00973-5}
}

\end{document}